\documentstyle[12pt,titlepage,epsf,cite]{article}
\newcommand{\nc}{\newcommand}

\nc{\bd}{
\begin{document}} 
\nc{\ed}{\end{document}} 
\nc{\be}{\begin{equation}}
\nc{\ee}{\end{equation}}
\nc{\ba}{\begin{eqnarray}}
\nc{\ea}{\end{eqnarray}}
\nc{\bfg}{\begin{figure}}
\nc{\efg}{\end{figure}}
\nc{\btb}{\begin{table}}
\nc{\etb}{\end{table}}

\nc{\nn}{\nonumber}
\nc{\ul}{\underline}
\nc{\ra}{\rightarrow}
\nc{\la}{\leftarrow}
\nc{\longra}{\longrightarrow}
\nc{\longla}{\longleftarrow}
\nc{\Ra}{\Rightarrow}
\nc{\La}{\Leftarrow}
\nc{\Longra}{\Longrightarrow}
\nc{\Longla}{\Longleftarrow}
\nc{\lra}{\leftrightarrow}
\nc{\Lra}{\Leftrightarrow}
\nc{\longlra}{\longleftrightarrow}
\nc{\Longlra}{\Longleftrightarrow}
\nc{\eps}{\epsilon }


\nc{\id}{{\sf 1\hspace*{-0.50ex}I\hspace*{1.0ex}}}
\nc{\C}{{\sf C\hspace*{-0.90ex}\rule{0.15ex}{1.5ex}\hspace*{0.9ex}}}
\nc{\N}{{\sf N\hspace*{-1.05ex}\rule{0.15ex}{1.1ex}\hspace*{1.0ex}}}
\nc{\Q}{{\sf Q\hspace*{-1.15ex}\rule{0.15ex}{1.5ex}\hspace*{1.1ex}}}
\nc{\R}{{\sf R\hspace*{-1.15ex}R\hspace*{0.2ex}}}
\nc{\Z}{{\sf Z\hspace*{-1.00ex}Z\hspace*{0.2ex}}}
                 

\nc{\xv}[1]{\langle #1 \rangle}  	
\nc{\half}{\frac{_1}{^2}}		
\nc{\sfrac}[2]{\frac{_{#1}}{^{#2}}}	
\nc{\nil}{\mbox{$\not\!o\ $}}		
\nc{\dslash}{\mbox{$\not{\hspace{-1.03mm}D}$}}        
\nc{\Dslash}{\mbox{$\not{\hspace{-1.03mm}D}$}}        

\nc{\tr}{\rm tr}
\nc{\asinh}{\rm asinh}
\nc{\acosh}{\rm acosh}
\nc{\atanh}{\rm atanh}


\nc{\Mo}{\stackrel{\circ}{M}}
\nc{\mo}{\stackrel{\circ}{m}}
\nc{\Wo}{\stackrel{\circ}{W}}
\nc{\gtk}{{\cal G}_{2k}}
\nc{\stk}{{\cal S}_{2k}}
\nc{\qtk}{{\cal Q}_{2k}}


\bd

\begin{titlepage}
\begin{flushright}
CU-TP-603 \\
July, 1993
\end{flushright}
\vspace*{\fill}
\vspace{.25in}

\begin{center}

{\Large\bf SEMI-ANALYTICAL SOLUTION OF THE $\phi^4$ THEORY
ON AN $F_4$ LATTICE}

\vspace{.25in}

Markus Klomfass
\footnote{\tt e-mail: klomfass@cuphyg.phys.columbia.edu} 

\vspace{.15in}

{\it Department of Physics, Columbia University, New York, NY 10027, USA}

\end{center}

\vspace{1.25in}
                
{\centering \bf ABSTRACT \\ }

Investigating
the cutoff dependence of the Higgs mass triviality bound, the
$\phi^4$ theory is formulated on an $F_4$ lattice which preserves
Lorentz invariance to a higher degree than the commonly used hypercubic
lattice.
I solve this model non-perturbatively by evaluating the high temperature
expansion through 13th order following the approach of L\"uscher and
Weisz. The results are continued across the transition line into the
broken phase by integrating the perturbative RG equations.
In the broken phase,
the renormalized coupling never exceeds 2/3 of the tree
level unitarity bound when $\Lambda/m_R \geq 2$.
The results confirm recent Monte Carlo data
and I obtain as an upper bound
for the Higgs mass
$m_R/f_\pi \leq 2.46 \pm 0.02_{\rm HTE} \pm 0.08_{\rm PT}$
at $\Lambda/m_R=2$.

\vspace*{\fill}
\end{titlepage}
  
\section{Introduction}

The relevance of the cutoff within the $\phi^4$ theory arises
because of triviality, i.e. the vanishing of the self-coupling 
in the continuum limit.
If the model were not trivial, all one would be interested in would be
the continuum limit, which of course is cutoff independent, and thus
this investigation into the cutoff dependence of the Higgs mass bound 
would be irrelevant.
Because of triviality, the cutoff has to be kept finite in order 
to retain a meaningful model at non-zero coupling.
In this way, the Standard Model has to be viewed as an effective 
low energy theory with a built in cutoff, 
embedded in some yet unknown full theory.
For self consistency, this cutoff should be at least 
a few times the Higgs mass, beyond the particle content of the
effective theory.
The closer the Higgs lies to the cutoff scale, the more important the
effects of the embedding theory (i.e. the cutoff effects) will be. 
Thus a search for a maximal Higgs mass
demands a careful investigation of the
cutoff dependence on the triviality bound.

To lowest order in the cutoff, the scalar sector is described accurately
by the usual $\phi^4$ Lagrangian.
The effect of the underlying full theory is to introduce corrections in
the form of (yet unknown) higher dimensional operators.
There are two dimension six operators parameterizing the leading order 
($\Lambda^{-2}$) cutoff effects.
The ordinary hypercubic lattice --- which is commonly used 
in Monte Carlo investigations of the Higgs bound ---
breaks Lorentz invariance already at order $\Lambda^{-2}$,
so the cutoff dependence on the lattice could look very different
than in the real world. The hypercubic lattice regularization 
could severely contaminate the physics.
The $F_4$ lattice, in contrast, accurately reproduces 
all physical ${\cal O}(\Lambda^{-2})$ corrections
and Lorentz breaking effects first occur at order $\Lambda^{-4}$.
Scalar fields can be implemented trivially on the $F_4$ lattice.
This idea was first  suggested by Neuberger \cite{neu:f4}.

Despite triviality, the cutoff may turn out to be so low that
the physical self-coupling could indeed be
strong (with respect to a tree level unitarity bound).
This calls for non-perturbative methods to investigate 
the triviality bound,
even though previous work on a hypercubic
lattice seems not to leave much room for a strongly coupled
scalar sector.
Besides Monte Carlo methods, where 
systematic errors arise because of the finite system size,
the strong coupling (or high-temperature) expansion has been worked out. 
Here results are obtained directly in the infinite volume limit, 
without any statistical errors.
The method, however, only works in the symmetric phase and relies
on the perturbative renormalization group to be continued into the
Goldstone phase. It is also quite cumbersome to apply 
the high-temperature expansion to
lattice actions that include  additional higher dimensional 
operators which modify the naive nearest neighbor interaction.

Monte Carlo simulations have been previously performed on an
$F_4$ lattice, but a strong coupling expansion 
of the model on this particular lattice has been lacking so far.
I will essentially follow the approach of L\"uscher and Weisz,
presented in a series of papers 
\cite{lw:lce,lw:trivbound1,lw:trivbound2,lw:trivbound3}
and extend the application to the $F_4$ lattice.
For completeness, I will repeat some of their techniques
and emphasize the adaptions necessary for the $F_4$ lattice.

I work within the Dashen-Neuberger approximation\cite{dashen}, i.e.
Gauge and Yukawa couplings are neglected. They are weak and 
may be included later via perturbation theory.
This approximation holds (i.e. captures the relevant physics)
if the top quark is well below 0.5~TeV (which seems to be 
favored by experiment).
In this limit the Higgs action turns into the action 
for an O(4)-symmetric $\phi^4$ model
\ba
	S_H &=& \int d^4x \left\{
	\frac{1}{2} \partial_{\mu} \phi \cdot \partial_{\mu} \phi
	+ \frac{\lambda}{8} (\phi^2-v^2)^2
	\right\} \;.
\ea
The two physical scales in the broken phase are set by the Higgs mass
and the pion decay constant $f_{\pi}=$~246~GeV.
Based on $f_\pi$, a physical self-coupling $g_R$ is defined as
\be
	\label{eq-ren:coupl}
	g_R= 3 \left(\frac{m_R}{f_\pi}\right)^2 \;.
\ee
Triviality of $\phi^4$ in four dimensions
is not rigorously proven but rests on perturbative
renormalization group (RG) results and Monte Carlo data
\cite{triv:anl,triv:review,lw:strong,triv:num,neu:higgs}.
Perturbative renormalization group leads to 
the relation
\be
	\label{eq-mr:scaling}
	m_R/\Lambda = C_1 (\beta_1 g_R)^{-\beta_2 / \beta^2_1}
	\exp\left(-\frac{1}{\beta_1 g_R}\right)
	\{1+ {\cal O} (g_R) \} \;\;\;\; {\rm as }\;\;\; 
	g_R \ra 0 \;,
\ee
with $C_1$ being a $\lambda$-dependent integration constant
that will be evaluated in the high-tempera\-ture expansion.
Removal of the cutoff $\Lambda$ will lead to a trivial theory.
Lowering the cutoff, however, $g_R$ is increasing accordingly.
But so are cutoff effects
and by putting a limit on the acceptable cutoff dependence 
in physical observables (for instance
pion--pion scattering at 90 degrees) one gets an upper bound on  
$g_R$ and hence on $m_R/f_\pi$ defined in a 
{\it lattice independent} way,
valid beyond the realms of lattice gauge theory.

The smaller the coefficient $C_1$ in (\ref{eq-mr:scaling})
(which depends on the higher dimensional operators in an improved action)
the less the impact a change in the cutoff will have on $m_R/f_\pi$. 
This in turn allows a higher mass bound. I refer to 
\cite{bha:f4_anl,bha:f4_num,hknv} for a detailed analysis.

This paper is organized as follows.
In the next section I will briefly review some perturbative results 
for the $\phi^4$ model on an $F_4$ lattice.
I will give scaling laws in the symmetric and in the 
broken phase. (Expressions up to three loops for 
the RG functions are listed in the appendix.)
These results were worked out by L\"uscher and Weisz 
\cite{lw:lce,lw:trivbound1,lw:trivbound2,lw:trivbound3}
and by Bhanot et al. \cite{bha:f4_anl,bha:f4_num}.
In the following section, 
I will give a detailed account of the high-temperature series 
in form of a linked cluster expansion. 
The method follows work by L\"uscher and Weisz
(henceforth referred to as LW) 
\cite{lw:lce}, but is adapted to the $F_4$ lattice.
The analysis of the expansion up to 13th order
is presented in the last section
(preliminary results were given in \cite{thesis,lat92}).
Results in the broken phase are obtained by propagating the
high-temperature expansion results using
the perturbative RG equations given in section~2.
I will show that the region where the cluster expansion converges
and the perturbative region do overlap on the $F_4$ lattice.
An estimate for the Higgs triviality bound is given and 
compared with Monte Carlo data.
Ultimately, this comparison serves as an indicator of 
the size of systematic errors. Furthermore, 
the $F_4$ result shows that Lorentz invariance breaking
effects are less than  10\%, indicating the robustness of the
triviality bound.

\section{Definitions, scaling laws and matching }

Many of the perturbative results 
required for the high-temperature expansion analysis
listed throughout this section are taken from 
\cite{lw:trivbound3} and \cite{bha:f4_anl}
and are reproduced here for completeness only.

The naive euclidean lattice action for the $\phi^4$ theory is
\be \label{eq-action}
      S = - 2\kappa \sum_{<x x'>} \phi(x) \phi(x')
    		+ \sum_{x} u\biggl(\phi(x)\biggr) \;,    
\ee
with a local O$(N)$ invariant potential
\be
      u(\phi) = \phi^2 + \lambda (\phi^2 -1)^2 \;.
	\label{eq-potnl}
\ee
The summation in (\ref{eq-action}) is over all nearest neighbor pairs $<x x'>$
and $\phi$ is a real $N$-component field 
$\phi^a(x),\; a=1,\ldots,N$, located on an
$F_d$ lattice. The parameter range is restricted to
$ \kappa \geq 0 $ and $\lambda \geq 0$.
The model is known to exist in two phases, separated by a second
order critical line $\kappa_c(\lambda)$.

The $F_d$ lattice in $d \geq 2$ dimensions is defined as the set
of points \linebreak 
$\{ x | x = \sum_{\mu} x_{\mu} e_{(\mu)}, x_{\mu} \in \Z ,\;
\sum_{\mu} x_{\mu}= {\rm even} \}$. Here $e_{(\mu)}$ is the 
euclidean unit vector in the $\mu$-direction. 
$F_d$ can be viewed as a hypercubic lattice $Z^d$ with its odd sites removed
and for $d=3$ it corresponds to an {\it fcc} lattice.
The euclidian distance between nearest neighbors is $\sqrt 2$ and
there are $2d(d-1)$ nearest neighbors per site.
The discrete rotation symmetry group is particularly large in four
dimensions and in this case the Lorentz invariance breaking term
in the kinetic energy part of the free propagator vanishes.
Invariance breaking terms first occur at order $\Lambda^{-4}$.

The generating functional $W$ for the connected correlation functions 
of $\phi$ and $\cal O$ 
in the presence of the external sources $J(x)$ and $K(x)$ is
\be
      \exp\biggl(W[J,K]\biggr) = 
	\frac{1}{Z[0,0]} \int \prod_{x,\alpha} d\phi^\alpha(x) \exp \left\{
      -S[\phi] + \sum_{x,\beta} J^\beta(x) \phi^\beta(x) 
	+ \sum_x K(x) {\cal O} (x) 
	\right\} \;,
\ee
where $\cal O$ is a composite operator defined as
${\cal O} (x) = 2 \sum_{<x x'>} \phi(x) \phi(x')$.
The connected correlation functions 
are then obtained by taking the derivatives 
of $W[J,K]$ with respect to the sources 
\ba
	\lefteqn{
      W^{(n,l)}(x_1,\ldots,x_n;y_1,\ldots,y_l)
	_{\alpha_1\ldots\alpha_n} =}  \nn \\
      & & \left. \frac{\delta^{n+l}}
	{\delta J_{\alpha_1}(x_1)\ldots \delta J_{\alpha_n}(x_n)
	\delta K(y_1)\ldots \delta K(y_l)}
        W[J,K] \right|_{J=K=0} \;.
\label{eq-znl}
\ea
The vertex functions $\Gamma^{(n,l)}$ are 
generated by the one particle irreducible functional $\Gamma[M,K]$
defined as the Legendre transform $\Gamma = W -\sum_x J(x) M(x)$.
Then
\ba
	\label{eq-gnl}
	\lefteqn{
      \Gamma^{(n,l)}(p_1,\ldots,p_n;q_1,\ldots,q_l)
	_{\alpha_1\ldots\alpha_n} = } \nn \\
	& &  \left. \frac{\delta^{n+l}}
	{\delta \tilde{M}(p_1)_{\alpha_1}\ldots 
	 \delta \tilde{M}(p_n)_{\alpha_n}
	\delta \tilde{K}(q_1)\ldots \delta \tilde{K}(q_l)}
      \Gamma[M,K] \right|_{M=K=0},
\ea
and  $M$ is the local magnetization, $M= \partial W/ \partial J$.
Because of O$(N)$ invariance, the coefficients $\Gamma^{(n,l)}$ 
vanish for odd $n$. 

The renormalization conditions imposed on the 
vertex functions in the symmetric phase at zero external momentum
define the wavefunction renormalization constant $Z_R$ and the
renormalized mass $m_R$, 
\be
	\label{eq-rc:g2}
	\Gamma^{(2,0)}(p,-p)_{\alpha \beta} =
	\delta_{\alpha \beta} \frac{1}{Z_R} \{ p^2 +m_R^2 \} 
	+ {\cal O} (p^4) \;\;, (p \ra 0) \;.
\ee
The renormalized coupling is fixed via
\be
	\label{eq-rc:g4}
	\Gamma^{(4,0)} (0,0,0,0)_{\alpha \beta \gamma \delta} =
	\frac{1}{3} C_4 (\alpha, \beta, \gamma ,\delta) 
	\frac{g_R}{Z_R^2} \;,
\ee
with $C_4$ being the totally symmetric O$(N)$ invariant tensor
$	C_4(\alpha, \beta, \gamma ,\delta) =
       \delta_{\alpha \beta}\delta_{\gamma \delta} +
       \delta_{\alpha \gamma}\delta_{\beta \delta} +
       \delta_{\alpha \delta}\delta_{\beta \gamma} $.
The field renormalization constant $ Z^{\cal O}_R$
associated with the composite operator ${\cal O}$ is given by
\be
	\Gamma^{(2,1)}(0,0;0)_{\alpha \beta} =
	\delta_{\alpha \beta} 
	\frac{1}{Z_R Z^{\cal O}_R} \;.
	\label{eq-zor}
\ee
The renormalized vertex functions are defined by 
\ba
	\Gamma^{(n,l)}_R &=& 0 \;\;\mbox{ for odd $n$ and for }
	 n=0,\; l\leq 1 \;, \nn \\
	\Gamma^{(n,l)}_R &=&
	Z^{n/2}_R Z^{{\cal O} \; l}_R \biggl \{ \Gamma^{(n,l)} 
	- \delta_{n 0} \delta_{l 2} \Gamma^{(0,2)} (0,0)
	\biggr \} \;\; \mbox{ otherwise } \;,
\label{eq-gamr}
\ea
and satisfy the following normalization conditions
\ba
	\Gamma^{(2,0)}_R (p,-p)_{\alpha \beta} &=&
	\delta_{\alpha \beta} \{ p^2 +m_R^2 \} 
	+ {\cal O} (p^4) \;, \;\; (p \ra 0) \;, \\
	\Gamma^{(4,0)}_R (0,0,0,0)_{\alpha \beta \gamma \delta} &=&
	\frac{1}{3} C_4(\alpha, \beta, \gamma ,\delta) 
	g_R \;, \\
	\Gamma^{(0,2)}_R (0,0) &=& 0 \;, \\
	\Gamma^{(2,1)}_R (0,0;0)_{\alpha \beta} &=& \delta_{\alpha \beta} \;.
\ea
From the lattice action, the vertex functions will be expressed in terms
of $\kappa$ and $\lambda$. 
Taking the derivative of the renormalized vertex functions
with respect to $\kappa$ at a fixed value of $\lambda$, one 
derives the {\em Callan-Symanzik} equation
\ba
	\left\{ m_R \frac{\partial}{\partial m_R}
	+ \beta \frac{\partial}{\partial g_R} - n \gamma -l \delta
	\right\} \Gamma^{(n,l)}_R  \hspace{4cm} \nn \\
        \hspace{2cm}
	= \epsilon m^2_R \left\{
	 \left . \Gamma^{(n,l+1)}_R \right |_{q_{l+1}=0}
	- \delta_{n 0} \delta_{l 2} \Gamma^{(0,3)}_R (0,0,0)
	\right\} \;.
\ea
The coefficients are defined by
\ba
	\label{eq-rg:beta}
	\beta (m_R,g_R) &=& m_R \frac{\partial g_R}{\partial \kappa}
	\left / \frac{\partial m_R}{\partial \kappa} \right.  \;, \\
	\label{eq-rg:gamma}
	\gamma (m_R,g_R) &=& \half m_R \frac{\partial \ln Z_R}
	{\partial \kappa}
	\left / \frac{\partial m_R}{\partial \kappa} \right. \;, \\
	\label{eq-rg:delta}
	\delta (m_R,g_R) &=& m_R \frac{\partial \ln Z^{\cal O}_R}
	{\partial \kappa}
	\left / \frac{\partial m_R}{\partial \kappa} \right. \;, \\
	\label{eq-rg:epsilon}
	\epsilon (m_R,g_R) &=& \left(
	m_R Z^{\cal O}_R \frac{\partial m_R}{\partial \kappa} 
	\right)^{-1} \;\;=\;\; 2(\gamma-1) \;.
\ea
Equivalently, one may write the above equations in the following way
\ba
	\label{eq-beta}
	m_R \frac{\partial g_R}{\partial m_R} &=& \beta \;, \\
	\label{eq-gamma}
	m_R \frac{\partial \ln Z_R}{\partial m_R} &=& 2 \gamma \;, \\
	\label{eq-delta}
	m_R \frac{\partial \ln Z^{\cal O}_R}{\partial m_R} &=& \delta \;, \\
	\label{eq-kappa}
	m_R \frac{\partial \kappa}{\partial m_R} &=& 
	m^2_R \eps Z^{\cal O}_R \;,
\ea
where all derivatives are taken at fixed $\lambda$ and $m_R$ is
the independent variable.  
At the transition, as $m_R \ra 0$, the RG functions
$\beta,\gamma,\delta$ and $\epsilon$ have finite limits.
The universal coefficients $\beta_{\nu},\gamma_{\nu},\delta_{\nu}$
and $\epsilon_{\nu}$
have been previously calculated up to three loops. They are given in
the appendix together with the non-universal scaling violating terms 
up to one loop.

Equation (\ref{eq-beta}) implies that, for a positive $\beta$-function,
as $m_R \ra 0$ the coupling also tends to zero as given in
eq.~(\ref{eq-mr:scaling}).
Similarly, scaling laws for $Z_R, Z^{\cal O}_R$ and $\kappa$
are derived from eqs.~(\ref{eq-gamma})--(\ref{eq-kappa}).
Specifically, one has
\ba
	\label{eq-zr:scaling}
	Z_R &=& C_2 \{1+ {\cal O} (g_R) \} \;, \\
	\label{eq-zor:scaling}
	Z^{\cal O}_R &=& C_3 (g_R)^{\delta_1 / \beta_1}
	\{1+ {\cal O} (g_R) \} \;, \\
	\label{eq-kappa:scaling}
	\kappa_c - \kappa &=& C_3 m^2_R 
	(g_R)^{\delta_1 / \beta_1}
	\{1+ {\cal O} (g_R) \} \;.
\ea
Switching from $m_R$ to a new independent variable $\tau=1-\kappa / \kappa_c$,
the equations are transformed to
\ba
	\label{eq-mrscal}
	m_R &\stackrel{\tau \ra 0}{\sim}& C_4 \;
	\tau^{1/2}|\ln \tau|^{\delta_1 / 2 \beta_1} \;, \\
	\label{eq-grscal}
	g_R &\stackrel{\tau \ra 0}{\sim}& 
	\frac{2}{\beta_1} |\ln \tau |^{-1} \;, \\
	\label{eq-zrscal}
	Z_R &\stackrel{\tau \ra 0}{\sim}& 
	C_2 \;, \\
	\label{eq-zorscal}
	Z^{\cal O}_R &\stackrel{\tau \ra 0}{\sim}& 
	C_5 |\ln \tau |^{-\delta_1 / \beta_1} \;.
\ea
At this point, note that the susceptibility (as defined in 
(\ref{eq-chi2}) below) is related to the renormalized
quantities via $\chi_2 = Z_R m^{-2}_R$,
resulting in the following scaling behavior
\ba
	\chi_2 &\stackrel{\tau \ra 0}{\sim}& 
	C_6 \; \tau^{-1}
	|\ln \tau |^{-\delta_1 / \beta_1} \;.
	\label{eq-chi2scal}
\ea
At small $\lambda$ however, the divergence of the 
susceptibility is dominated by the $\tau^{-1}$ term, unless one
is very close to the critical line. This can be seen as follows.
Keeping terms 
proportional to the initial value $g_0=6\lambda/(d-1)^2\kappa^2$
when integrating the RG equations, one
obtains
\ba
	g_R &\stackrel{\tau \ra 0}{\sim}& 
	g_0 \left\{ 1+ \frac{\beta_1 g_0}{2}
	| \ln \tau | \right\}^{-1} \;,
\ea
and thus the improved scaling law for the susceptibility
\ba
	\chi_2 &\stackrel{\tau \ra 0}{\sim}& 
	C_7 \; \tau^{-1}
	\biggl (
	1 - f(\lambda) |\ln \tau | \biggr )^{-\delta_1 / \beta_1} \;.
	\label{eq-chi2scali}
\ea
Here I have defined
\be
	f(\lambda) = \frac{d^2}{4 \pi^2} \frac{N+8}{N+2} \bar{\lambda}
	+ {\cal O}(\bar{\lambda}^2) \;,
	\label{eq-flambda}
\ee
where $\bar{\lambda}=(N+2)\lambda + {\cal O}(\lambda^2)$ 
(see (\ref{eq-lambar}) for a
proper definition) and I used
the fact that $\kappa$ is close to
$\kappa_c = 1/2d(d-1) + {\cal O} (\lambda)$.
All the other scaling laws are modified accordingly.
The improvement indicates that the logarithmic divergence is
suppressed for small $\lambda$.

In the broken symmetry phase, the $N$-th component of
$\phi$ is given a vacuum expectation value $v$
 while the first $N-1$ components remain massless. 
The generating functionals are defined in the limit 
of a vanishing external magnetic field $h$.
The vertex functions $\Gamma^{(n,l)}$ are again given by (\ref{eq-gnl}).
However, in the broken phase they are only O($N-1$) invariant.
The perturbative expansion of the vertex functions 
in the broken phase is defined around 
$\phi_\alpha=\delta_{\alpha N}s_{\rm min}$, where $s_{\rm min}$ 
is one of the two degenerate absolute minima of the lattice action.

The renormalization conditions in the broken phase are 
imposed (following \cite{bha:f4_anl}) as
\ba
	\Gamma^{(2,0)}_R (p,-p)_{\alpha \beta} \; = \;
	\delta_{\alpha \beta} \;p^2  
	\!\!\!&+&\!\!\! {\cal O} (p^4) \;, \;\; (p \ra 0) \;, \\
	\label{eq-rc:mr}
	\mbox{Re}\biggl\{
	\left. \Gamma^{(2,0)}(p,-p)_{NN} \right|_{p=(im_R,0,0,0)}
	\biggr\} &=& 0 \;, \\
	\mbox{Re}\biggl\{
	\left. \Gamma^{(1,1)}(p;-p)_{N} \right|_{p=(im_R,0,0,0)}
	\biggr\} &=& \frac{v}{Z_R Z^{\cal O}_R} \;.
\ea
The pion decay constant is just the vacuum expectation value of the
renormalized $\sigma$-field
$f_\pi = (Z_R)^{-\half} v$.
Now the renormalized self-coupling is defined as 
in eq. (\ref{eq-ren:coupl}) and 
renormalized vertex functions are introduced via
\ba
	\Gamma^{(n,l)}_R &=& 0 \;\;\mbox{ for } n=0,\; l\leq 1 \;,  \\
	\Gamma^{(n,l)}_R &=& 
	Z^{n/2}_R Z^{{\cal O} \; l}_R \biggl \{ \Gamma^{(n,l)} 
	- \left. \delta_{n 0} \delta_{l 2} \Gamma^{(0,2)} (p,-p)
	\right|_{p=(im_R,0,0,0)}
	\biggr \} \;\; \mbox{ otherwise } \nn \;.
\ea
The Callan-Symanzik equation in the broken phase reads
\ba
	\left\{ m_R \frac{\partial}{\partial m_R}
	+ \beta \frac{\partial}{\partial g_R} - n \gamma -l \delta
	\right\} \Gamma^{(n,l)}_R  \hspace{4cm} \nn \\
        \hspace{1cm}
	= \vartheta m_R
	 \left. \Gamma^{(n+1,l)}_R \right |_{p_{n+1}=0,\,\alpha_{n+1}=N}
	+ \eps m^2_R
	 \left. \Gamma^{(n,l+1)}_R \right |_{q_{l+1}=0}
\ea
The RG functions are defined as in
eqs.~(\ref{eq-rg:beta})---(\ref{eq-rg:epsilon}). 
However, the relation between $\eps$ and $\gamma$ is not valid 
in the broken phase, and
--- unfortunately --- in the case of $N=4$ there exists no simple relation
between $\eps$ and the other RG functions.
The function $\vartheta$ is defined as
\ba
	\label{eq-rg:theta}
	\vartheta(m_R,g_R) &=& Z_R^{-1/2} 
	\frac{\partial v}{\partial \kappa} \left/
	\frac{\partial m_R}{\partial \kappa} \right. \;,
\ea
and it is related to the other functions through
\ba
	\vartheta &=& \left(1+\gamma -\frac{1}{2g_R}\beta
	\right) \sqrt{\frac{3}{g_R}} \;.
\ea 
The universal coefficients $\beta_{\nu},\gamma_{\nu},\delta_{\nu}$
and $\epsilon_\nu$ plus the scaling violating terms
are listed in the appendix.

The scaling laws for $m_R$, $Z_R$, and $Z^{\cal O}_R$ --- given in
(\ref{eq-mr:scaling}), (\ref{eq-zr:scaling}) and (\ref{eq-zor:scaling}) --- 
are exactly the same, except with integration constants $C'_i$ instead
of $C_i$. The only change is that
in the scaling law (\ref{eq-kappa:scaling})
for $\kappa$ appears an additional factor of $-1/2$
\ba
	\kappa - \kappa_c &=& \frac{1}{2} C'_3 m^2_R 
	(g_R)^{\delta_1 / \beta_1}
	\{1+ {\cal O} (g_R) \} \;.
\ea
As in the symmetric phase, the scaling law for $m_R$ implies that
the renormalized coupling tends to zero
when approaching the critical line from above.

To connect the theory in the symmetric with the theory in the broken phase,
one has to establish a relationship between the integration constants
$C_i$ and $C'_i$ on both sides of the transition.
This is accomplished by reconstructing the massive 
renormalized vertex functions in both
phases from the critical renormalized vertex functions
by mass perturbation theory.
The answer can be obtained from a one-loop calculation. One finds
\ba 
	\label{eq:match}
	C'_1(\lambda) &=& C_1(\lambda) \; \exp \left\{
	\frac{2N+17-3\sqrt{3}\pi}{2N+16}
	\right\} \;, 
\ea
and $C'_i = C_i$ for $i=2,3$.

Finally, the tree level unitarity bound in the symmetric phase 
from the $S$-wave phase shift for elastic scattering 
amounts to $g_R<29$ for $N=4$.
An almost equal bound holds in the broken phase \cite{lw:trivbound3}.

\section{The linked cluster expansion}

As a powerful non-perturbative semi-analytical tool,
the high-temperature expansion 
serves as an alternative to Monte Carlo methods.
It has been widely used and for details I refer to \cite{lw:lce,wortis}.
The high-temperature expansion can be systematically
organized in terms of graphs consisting of vertices and
connecting lines. 
This method is referred to as {\em linked cluster expansion}.
The number of lines in a graph corresponds to the 
power of the hopping parameter $\kappa$ in the high-temperature expansion.
I compute susceptibilities up to 13th order in $\kappa$
with a total of about 400~000 contributing graphs. 
The number of contributing graphs depends on the lattice
structure as not all graphs can be embedded on all lattices.
The whole procedure is computerized as outlined in \cite{lw:lce}
and is easily adaptable for different potentials and 
lattice structures.

It is sufficient to define the 
following susceptibilities in terms of the connected
correlation functions $W^{(2,0)}$ and $W^{(4,0)}$
\ba
      \chi_2 &=& \sfrac{1}{N} \sum_{x,a} \xv{\phi^a(x)\phi^a(0)}^{\rm conn}\;, 
	\label{eq-chi2}\\
      \mu_2 &=& \sfrac{1}{N} \sum_{x,a} x^2 \xv{\phi^a(x)\phi^a(0)}^{\rm conn}\;,
	\label{eq-mu2} \\
      \chi_4 &=& \sfrac{1}{N^2} \sum_{\stackrel{x,y,z}{a,b}}
       \xv{\phi^a(x)\phi^a(y)\phi^b(z)\phi^b(0)}^{\rm conn} 
	\label{eq-chi4}\;,
\ea
where the spatial summation produces zero momentum operators.
From the expansion of the 
susceptibilities an expansion for the quantities of interest, particularly
the mass $m_R$, the coupling $g_R$, and the wavefunction renormalization
constants $Z_R$ and $Z^{\cal O}_R$ can be calculated 
in principal anywhere in the symmetric phase via
\ba
	\label{eq-mreval}
      m_R   &=&	\left( 2d \frac{\chi_2}{2\mu_2} \right)^{1/2} \;, \\
	\label{eq-greval}
      g_R   &=& - (2d)^2 \; \frac{\chi_4}{2\mu_2^2} \;, \\
	\label{eq-zreval}
      Z_R   &=& 2d \; \frac{\chi_2^2}{\mu_2} \;, \\
	\label{eq-zoreval}
	Z^{\cal O}_R &=& \frac{\mu_2}{d} 
	\left( \frac{\partial \chi_2}{\partial \kappa} \right)^{-1} \;.
\ea
These relations can easily be derived from the renormalization conditions
imposed on the vertex functions in the symmetric phase.
To extend the solution into the broken phase one has to resort
to other means, e.g. perturbative RG techniques. 

Care has to be taken in defining a proper Fourier transform on
the underlying lattice structure.
The particular form of the Brillouin zone on the 
$F_4$ lattice results in additional factors of 2 in
the above relations as compared to the hypercubic case.

I will now introduce the necessary ingredients for the 
cluster expansion, starting with some basic definitions.
A graph $G$ is defined by a set $({\cal V, L ,E})$, where
${\cal V}$ represents the {\it vertices}, 
${\cal L}$ the {\it internal lines} connecting the vertices and 
${\cal E}$ the number of {\it external lines} attached to each vertex.
Let $V$ be the total number of vertices, 
$L$ the number of internal lines or {\it order} of the graph,
$E$ the total number of external lines.
Let $N(v)=I(v)+E(v)$ be the total number of lines, internal plus external,
attached to a vertex $v$.
Two graphs $G$ and $G'$ are called {\it topologically equivalent}, if
$G'$ can be transformed into $G$ by simply reordering its vertices and
$[G]$ is the corresponding equivalence class of graphs.
In the linked cluster expansion, the particular order of the
vertices is irrelevant and therefore the expansion is in terms
of equivalent classes rather than individual graphs.
Care has to be taken to avoid overcounting.

The  {\it distance} between two vertices $d(v_i,v_j)$ is the length of the
shortest path between $v_i$ and $v_j$. 
A  {\it loop} is a closed path.
Generally, a graph may contain vertices that are not connected
by a path to some (or all) other vertices. 
A graph $G$ is said to be  {\it connected}
if there is at least one path between any two vertices.
A connected graph $G$ is said to be  {\it non-separable}
if for all vertices $v$ the graph $G$
remains connected after removal of $v$ and all
lines emanating from it. 
A graph is called {\it simple} if any two vertices are connected 
by at most one line and there are no self-loops 
(i.e. no lines starting and ending at the same vertex).

The linked cluster expansion is based solely on graphs without self-loops, 
and therefore contributing graphs may be represented 
by a $V \times V$ symmetric matrix.
The entries $A(i,j)$ of this so-called {\it incidence matrix} 
give the number of internal lines from $v_i$ to $v_j$ for $i\neq j$,
and $A(i,i)= E(v_i)$ otherwise.
The incidence matrices of topologically equivalent graphs are related 
through a permutation of the vertex labelling $(1,\ldots,V)$.
To eliminate multiple occurrences of equivalent graphs in the
expansion,
one has to define a certain canonical ordering of the vertices
which must  depend only on the topology
of the graph under consideration. 
As a result, every equivalence class of graphs is represented by
a single incidence matrix $A(i,j)$.

With each graph in the expansion one associates 
certain characteristic numbers which may depend on three 
sources: the topology of the graph, the underlying lattice
structure or the local potential in the action.

The potential $u(\phi)$ only enters through the following weight
factors. 
For a graph $G$ with a set of vertices ${\cal V} = (v_1,\ldots,v_V)$
it is defined as 
\be
      \Wo(G) = \prod_{v \in {\cal V}} \mo^{\rm conn}_{N(v)} \;,
	\label{eq:weight}
\ee
where the $\mo^{\rm conn}_{2k}$ are the so-called connected moments.
They can be calculated from the potential as follows.

The Taylor expansion of the free energy $W[J,K;\kappa]$ 
at zero coupling where all spins are uncorrelated 
and essentially randomly distributed can be written in terms of
the  so-called {\em cumulants} $\Mo^{a_1 \ldots a_k}_k(x_1)$ 
defined via
\ba \label{eq-mo}
   \left. \frac{\partial^k W}
   {\partial J^{a_1}(x_1)\ldots\partial J^{a_k}(x_k)} \right|_{\kappa=0} 
    &=&  \delta(x_1,\ldots,x_k) \Mo^{a_1 \ldots a_k}_k(x_1) \;.
\ea
For zero external field, only the even correlation functions
$W^{(2k,0)}$ survive in a symmetric potential
and  the cumulants take a simple form
\ba
      \left. \Mo^{a_1\ldots a_{2k}}_{2k} (x_1)
      \right|_{J=K=0} 
      &=& \left. \xv{\phi^{a_1}(x_1)\ldots\phi^{a_{2k}}(x_{2k})}^{\rm conn}
	\right|_{\kappa=0} \nn \\
      &=& \delta(x_1,\ldots,x_{2k}) 
	\xv{\phi^{a_1}(x_1)\ldots\phi^{a_{2k}}(x_1)}^{\rm conn} \;,
\ea
where the connected correlation functions are evaluated at one point
in space-time and may be denoted by
$\xv{\phi^{a_1}\ldots\phi^{a_{2k}}}_1^{\rm conn}$.
From O$(N)$ invariance it follows that
\ba  
      \xv{\phi^{a_1}\ldots\phi^{a_{2k}}}_1^{\rm conn}
      &=& C_{2k}(a_1,\ldots,a_{2k}) \mo^{\rm conn}_{2k} \delta(x_1,\ldots,x_{2k}) 
      \;, \label{eq-conmom}
\ea
where $\mo^{\rm conn}_{2k}$ are the {\em connected moments}
appearing in (\ref{eq:weight}).
The $C_{2k}(a_1,\ldots,a_{2k})$ are O$(N)$ invariant tensors.
The connected moments can easily be calculated from the
(not necessarily connected) moments, which are defined in terms of 
the correlation functions
\be \label{eq-mom}
      \xv{\phi^{a_1}\ldots\phi^{a_{2k}}}_1 = C_{2k}(a_1,\ldots,a_{2k})
      \mo_{2k} \;.
\ee
Using the proper normalization condition for the O$(N)$ tensors
one derives
\ba
      \mo_{2k} &=& \frac{\Gamma(\half N)}{2^k \Gamma(\half N + k)} 
      		    \frac{J_{N-1+2k}}{J_{N-1}} \;, 
	\label{eq-smo}
\ea
with
\be
      J_{l} = \int_{0}^{\infty} dr \; r^l \exp \biggl(-u(r)\biggr) \;\;.
	\label{eq-jint}
\ee
In the limit $\lambda \rightarrow \infty$, the moments can be written 
as a ratio of polynomials in $N$. One has
$\exp(-u(\phi)) \propto \delta(\phi^2-1)$ which results
in $J_l = const.$ and thus
\ba
      \mo_{2k} &=& \frac{1}{(N+2k-2)(N+2k-4)\ldots(N+2)N} \;\;.
\ea
From the moments the connected moments are generated via the relation
\be
      \sum_{k=1}^{\infty} \mo^{\rm conn}_{2k} \frac{z^k}{k!} =
      \ln \left\{1 + \sum_{k=1}^{\infty} \mo_{2k} \frac{z^k}{k!}
      \right\} \;\;.
\ee
This concludes the calculation of the weight factors.

Next, with every graph $G$, one defines an internal
symmetry number $S(G)$ as 
\be
      S(G)= S_{\pi}(A) \prod_{i<j} A(i,j)! \;\;,
	\label{eq-sg}
\ee
where $S_{\pi}(A)$ is the number of permutations that leave
the incidence matrix invariant and
the product represents the internal line multiplicity factor.
A modified symmetry number $S_E(G)$ is defined which
additionally takes the external line multiplicity into account
by setting $i\leq j$ in the above expression.

The underlying lattice structure enters through
the lattice embedding number $I(G)$ which counts the number of 
embeddings of $G$ on the lattice under these conditions:
the first vertex in $G$ is fixed at the origin of the lattice and 
all other vertices are then distributed in such a way
that any two vertices with
distance $1$ are placed on nearest neighbor lattice sites.
Different vertices may be placed on the same site and different lines
may occupy the same lattice link. 
Not all graphs are possible on
a certain lattice structure. For example, graphs containing odd loops
cannot be embedded on a hypercubic lattice.
Note that for the susceptibility $\mu_2$, one has to introduce a modified
lattice embedding number $I_2(G)$ which takes into account the
additional factors of $x^2$. 

Finally, one defines a symmetry number associated with the 
internal O$(N)$ symmetry.
For a graph $G$ with a set of internal lines ${\cal L}= (l_1,\ldots,l_L)$ 
one writes down an O$(N)$ index $a_l \in \{1,2,\ldots,N\}$ 
for every line $l$. Let 
$ \{ i_1,i_2,\ldots,i_{I(v_i)}\}$
be the set of indices labelling the internal lines which 
terminate in vertex $v_i$. Then the O$(N)$ factor for $G$ is given by
\be
      C(G) = \sum_{a_1,\ldots,a_L=1}^{N} \; \prod_{i=1}^{V}
      C_{N(v_i)}
      (\underbrace{1,1,\ldots,1}_{E(v_i) \; \rm{times}}
      ,a_{i_1},a_{i_2},\ldots,a_{i_{I(v_i)}}) \;.
\ee
$C(G)$ is a polynomial in $N$ with maximum degree $\half L$.

Having defined all the basic ingredients, one may now
write down the unrenormalized expansion for the susceptibilities
based on all equivalence classes of 
connected graphs without self-loops and with
$N(v)$ even for all $v \in {\cal V}$ and $E=2k$.
These graphs are called $\gtk$ graphs and the expansion reads
\ba
      \chi_2 &=& \sum_{G \in {\cal G}_2}^{} (2\kappa)^L I(G)C(G)\Wo(G)
      \frac{E!}{S_E(G)} \;\;, \\
      \mu_2  &=& \sum_{G \in {\cal G}_2}^{} (2\kappa)^L I_2(G)C(G)\Wo(G)
      \frac{E!}{S_E(G)} \;\;, \\
      \chi_4 &=& \sum_{G \in {\cal G}_4}^{} (2\kappa)^L I(G)C(G)\Wo(G)
      \frac{E!}{S_E(G)} \;\;.
\ea
It can be recast in the form
\ba
	\label{eq-chi2:form}
      \chi_2 &=& \sum_{L=0}^{\infty}
      \chi_2^{(L)} \frac{(2\kappa)^L}{L!}  \;\;, \\
	\label{eq-mu2:form}
      \mu_2  &=& \sum_{L=0}^{\infty}
      \mu_2^{(L)} \frac{(2\kappa)^L}{L!}  \;\;, \\
	\label{eq-chi4:form}
      \chi_4 &=& \sum_{L=0}^{\infty}
      \chi_4^{(L)} \frac{(2\kappa)^L}{L!}  \;\;,
\ea
where the coefficients 
$\chi_2^{(L)}$~,~$\mu_2^{(L)}$ and $\chi_4^{(L)}$ 
are of the form
\be
	\label{eq-coef:form}
      p[N] \prod_{k=1}^{\infty}(\mo^{\rm conn}_{2k})^{n_k} \;,
\ee
with $\sum_{k=1}^{\infty} k\; n_k  = L + \half E $.
The $p[N]$ are polynomials in $N$ of order not higher than $\half L$.

In order to reduce the total number of graphs, 
two levels of renormalization are carried out.
A reduction in the number of contributing diagrams in the
expansion is paid for with an increase in algebraic complexity.

The first step is to eliminate one-particle reducible graphs.
A graph $G \in \gtk$ is called {\it one-particle irreducible}, or 1PI, if it 
cannot be broken into two disconnected parts by cutting a single internal
line. The set of 1PI graphs is denoted by $\gtk^{1PI}$.
The expansion over $\gtk$ can be reconstructed via
\ba
      \chi_2 &=& \chi^{1PI}_2 (1-\epsilon\chi^{1PI}_2)^{-1} \;\;, \\
      \mu_2  &=& \biggl[\mu^{1PI}_2 + \epsilon (\chi^{1PI}_2)^2\biggr] 
      (1-\epsilon\chi^{1PI}_2)^{-2} \;\;, \\
      \chi_4 &=& \chi^{1PI}_4 (1-\epsilon\chi^{1PI}_2)^{-4} \;\;, 
\ea
where $\chi^{1PI}_2$ , $\mu^{1PI}_2$ and $\chi^{1PI}_4$ stand for
the 1PI part of the corresponding reducible quantities
and $\epsilon = q 2 \kappa $. Here $q$ is the coordination number
of the lattice.

In a second step, one throws out one-vertex reducible graphs.
A graph $G \in \gtk^{1PI}$ is called  {\it one-vertex irreducible}, or 1VI, if 
--- after removal of any vertex and all the lines emanating from it ---
each remaining connected piece of $G$ contains at least one external line.
The set of 1VI graphs is denoted by $\stk$.
Given the set $\stk$, the set $\gtk^{1PI}$ can be reconstructed by
attaching $\stk$ graphs repeatedly to all $G \in \stk$ 
in a certain manner which I will not describe here.
Furthermore, the connected moments $\mo^{\rm conn}_{2k}$ 
for the $\stk$ graphs have to be redefined as
\be
      m_{2k} = \frac{1}{(2k-1)!!} \sum_{G \in {\cal P}_{2k}}
      (2\kappa)^L I(G)C(G)\Wo(G)\frac{1}{S(G)} \;.
	\label{eq-renmom}
\ee
The $m_{2k}$ are called {\em renormalized moments} and are evaluated over
the set ${\cal P}_{2k}$, which is given by all graphs $G \in \gtk$ with the
additional constraint that all $2k$ external lines are attached to a single
vertex.
To lowest order one has $m_{2k} = \mo^{\rm conn}_{2k} + O(\kappa^2)$.
The structure for the moments is the same as for the susceptibilities
above. One finally has
\ba
	\label{eq-1pichi2}
      \chi^{1PI}_2 &=& \sum_{G \in {\cal S}_2} (2\kappa)^L I(G)C(G)W(G)
      \frac{E!}{S_E(G)} \;\;,\\
	\label{eq-1pimu2}
      \mu^{1PI}_2 &=& \sum_{G \in {\cal S}_2} (2\kappa)^L I_2(G)C(G)W(G)
      \frac{E!}{S_E(G)} \;\;,\\
	\label{eq-1pichi4}
      \chi^{1PI}_4 &=& \sum_{G \in {\cal S}_4} (2\kappa)^L I(G)C(G)W(G)
      \frac{E!}{S_E(G)} \;\;.
\ea

I will not go into details of the coding; the interested reader is
referred to \cite{lw:lce} and \cite{thesis}. 
I will only give a brief outline and stress the differences 
to the hypercubic case.

To construct the set of $\stk$ graphs required for 
the 1PI susceptibilities,
I start out with a set of 'simple' graphs which will
be expanded into the full set by a couple of basic operations.
Let ${\cal C}_1$ be the set of simple and connected graphs without
external lines which are 1PI and non-separable.
As external lines are absent, non-separability is equivalent to 
one-vertex irreducibility.
I denote the subset of ${\cal C}_1$ graphs with $L$ internal lines
by ${\cal C}_1(L)$.
All the  ${\cal C}_1(L)$ graphs can now be constructed
by repeatedly applying two simple operations to the
triangular graph, $C_1(3)$, represented by
\[
	\left(
	\begin{array}{ccc}
	0 & 1 & 1 \\ 
	1 & 0 & 1 \\ 
	1 & 1 & 0    
	\end{array}
	\right) \;\;. 
\]
(The zeroth order graph has to be added by hand.)
All higher order graphs can then be generated 
\begin{itemize}
\item  by adding a single line
between any two vertices with $A(i,j)=0$, and 
\item  by replacing a line by a bridge:
a new graph is generated by removal of a line, say
from $v_i$ to $v_j$ and 
addition of  a new vertex $w$ which is
connected to both, $v_i$ and $v_j$, by a single line each.
The number of lines and the number of vertices is thus increased by
one. 
\end{itemize}

It is obvious that the resulting graphs are still 
simple and connected.
However, the same graph may be constructed more than once. On the
other hand no graph in ${\cal C}_1$  is left out, as any graph
may always be reduced to the basic third
order graph by simply applying the inverse operations.
On a hypercubic lattice,
graphs with odd loops are ruled out and  one would therefore
start with the fourth order square graph.
In this case it is convenient to modify the rules of construction
such that graphs with zero lattice embedding number are not 
constructed at all.
The number of
${\cal C}_1(L)$ graphs up to 13th order is given in table~1 for
the general (i.e. $F_4$) case and for the hypercubic lattice.

\btb[p]
\caption[]
{Number of graphs contributing to ${\cal C}_1$,
${\cal S}_2$, ${\cal S}_4$ and $\qtk$.
The second row gives the number of graphs that can be embedded
on a hypercubic lattice. The numbers for $\qtk$ are the 
sum of all contributing graphs, $k=1,\ldots,\infty$, at the given order.}
\bigskip\centering{\footnotesize
\begin{tabular}{*{5}{r}}
\hline \hline 
$L$ & ${\cal C}_1$ & ${\cal S}_2$ & ${\cal S}_4$ & $\qtk$ \\
\hline
  0&	1&      1&      1&    	0	\\
   &	1&	1&	1&	0	\\
\hline
  1&     0&	0&      0&    	0	\\
   &	0&	0&	0&	0	\\
\hline
  2&     0&	0&      1&    	1	\\
   &	0&	0&	1&	1	\\
\hline
  3&     1&   	1&      2&    	1	\\
   &	0&	1&	1&	0	\\
\hline
  4&     1&   	1&      6&    	2	\\
   &	1&	0&	4&	2	\\
\hline
  5&     2&   	4&     13&    	3	\\
   &	0&	2&	4&	0	\\
\hline
  6&     4&   	8&     44&    	7	\\
   &	2&	3& 	20&	4	\\
\hline
  7&     7&  	22&    120&    	16	\\
   &	1&	8&	27&	0	\\
\hline
  8&     16& 	 57&    416&   	41	\\
   &	4&	9&	117&	19	\\
\hline
  9&     42& 	184&   1364&   	106	\\
   &	5&	40&	214&	0	\\
\hline
 10&     111& 	559&   4935&   	309	\\
   &	14&	68&	815&	80	\\
\hline
 11&     331& 	1910&  17952&  	932	\\
   &	20&	247&	1830&	0	\\
\hline
 12&     1098& 	6580&  68774&  	2995	\\
   &	65&	470&	6721&	509	\\
\hline
13 &	3829&	24046&	268524&	9972    \\
   &     124&	1779&	17028&	0	\\
\hline \hline 
\end{tabular}}
\etb

To construct the set ${\cal S}_{2k}$ from the set ${\cal C}_1$,
two additional conditions have to be satisfied. The total number of
lines attached to each vertex $N(v)$ has to be even, and the number of
external lines has to be $2k$.
For the graph to be 1VI, at least two vertices must have external
lines.
Besides adding external and multiple internal lines to fulfill
these conditions,
the resulting graphs need to be coupled to generate the 
complete set of ${\cal S}_{2k}$ graphs which are not necessarily
non-separable but only 1VI. Details are given in \cite{thesis}.
The number of ${\cal S}_{2k}$ for $k=1,2$ is also shown in table~1.

The total number of graphs generated up to 13th order is 
about 400,000. With 32 bit precision, this would require
$\half \cdot 13\cdot 14\cdot 4\cdot10^5\cdot 4$~bytes  = 145.6~Mb
of storage on the computer,
However, the incidence matrices can be stored in compressed form.
Each matrix element requires a maximum of 4 bits at 13th order.
Thus 12 4-byte integers are sufficient to represent a single
matrix. A subset of graphs is simple and can be represented
by just 3 integers. 
The total storage is therefore less than 19.2Mb.

\bfg[p]
\epsfxsize=15cm
\hspace{1. truecm}
\centering{\epsfbox{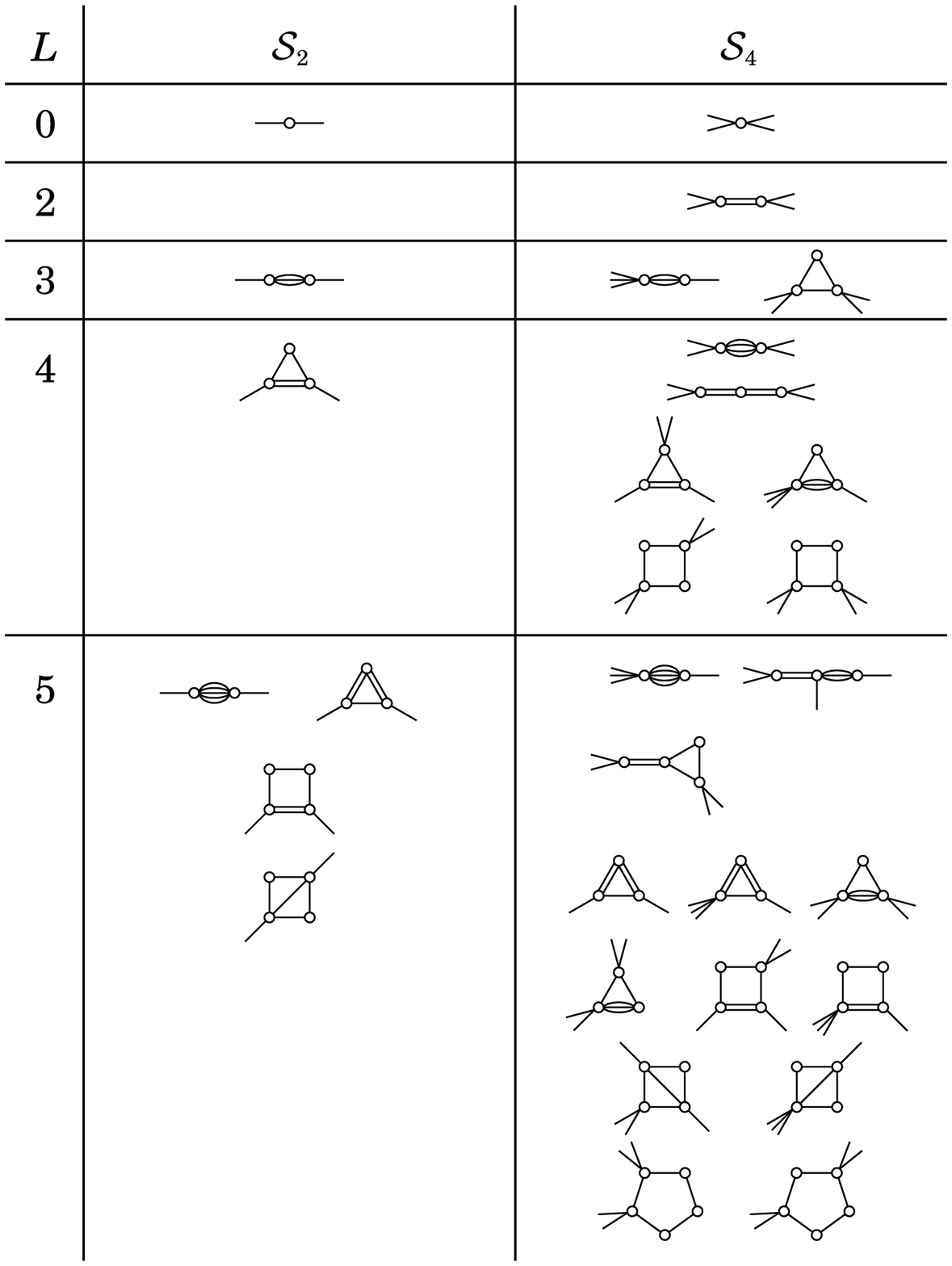}}
\hspace{1. truecm}
\caption[]
{The contributions to  ${\cal S}_2$ and ${\cal S}_4$
up to 5th order.}
\efg

The internal symmetry numbers and 
the $O(N)$ symmetry numbers are independent of the lattice
structure and are calculated just as in the hypercubic case.
For a detailed description of the  procedure see \cite{lw:lce}.
The lattice embedding numbers are more involved
and depend on the choice of lattice.
They are calculated for a subset of so-called reduced graphs
in order to minimize the calculational effort.
There is a total of $6180$ of these graphs up to 13th order,
as compared with a total of 396,140 $\stk$ graphs ($k=1,2$).
The evaluation of the embedding number involves 
the number of random walks ${\cal N}_l[x]$
from the lattice origin to $x$ in $l$ steps.
Instead of using a combinatorial approach to calculate 
these numbers, they are computed by actually carrying out the 
random walks on the lattice and creating a look-up table. This method 
can be adapted for different lattice structures without much effort.
For the modified lattice embedding numbers each
contribution is multiplied by an additional factor of 
$(x(i)-x(j))^2$, where $v_i$ and $v_j$
are the vertices with external lines. 
Factors of $\sqrt 2$ enter on the $F_4$ lattice due to the 
length of basis vectors.

The renormalized moments are evaluated 
as in (\ref{eq-renmom}) over the set ${\cal P}_{2k}$
which unfortunately is one-vertex reducible and therefore quite large.
By introducing yet another set of 1VI graphs
and new moments associated with these, it is possible 
to  compute the renormalized moments  order
by order in terms of the new moments in a recursive manner.
Consider a graph $G \in {\cal G}_2$ with $E(w)=2$
and $I(w)=2k$ for some vertex $w$.
Now construct a new graph $G'$ by removing the vertex $w$ plus both its
external lines. Note that $G'$ is not necessarily connected nor 1PI or
1VI. One then defines
\be
      \qtk = \left\{
      G \in {\cal G}_2 | G' \mbox{ connected and 1VI } 
      \right\} \nn \;.
\ee
$G'$ does not need to be 1PI. 
By construction, all graphs in ${\cal Q}_{2k}(L)$
will satisfy $L \geq 2k $.
Observe that for $G'$ to be one-vertex
irreducible, all spikes must end in an external line.
All external lines must have been connected to the
vertex $w$ before its removal. 
Therefore, $G$ will consist of loops only. 
This in turn means that all $\qtk$ graphs will in fact be 1PI.
For a hypercubic lattice,
all loops have an even number of lines and since $N(v)$ is even, too,
the graphs $G$, as well as $G'$
must be of even order.
This condition does not necessarily hold on other lattices, though, and
in particular it is not true on an $F_4$ lattice.
Finally, the new moments are evaluated in terms of the ${\cal Q}_{2k}(L)$
graphs and these are related to the renormalized moments in a
recursive manner.
Again, the full procedure is described in detail in \cite{lw:lce}.

With the renormalized moments, I have all the necessary
components to evaluate the one particle irreducible 
parts of the susceptibilities
as given in eqs.~(\ref{eq-1pichi2})---(\ref{eq-1pichi4}).
After a simple re-expansion of the series, I obtain the final answer in the
form of eqs.~(\ref{eq-chi2:form})---(\ref{eq-chi4:form}).
Limited by computational resources, I carry out the expansion 
up to 13th order. 
On a hypercubic lattice, where the number of graphs is 
significantly fewer, L\"uscher and Weisz were able to go to
14th order.

The total number of contributing terms in the expansion for $L \leq 13$
on the $F_4$ lattice is 
387 for $\chi_2$, 357 for $\mu_2$, and 562 for $\chi_4$.
The coefficients are given in tables~2---4 for the non-linear limit. 
For general $\lambda$, the coefficients are defined in terms of polynomials
$p[N]$ multiplied by powers of the connected moments, see
eq.~(\ref{eq-coef:form}). These numbers can be obtained 
from the author by electronic mail. 

As a check on the computer program, 
I reproduce the high-temperature expansion up
to 14th order on a hypercubic lattice. I am in full agreement
with L\"uscher and Weisz \cite{lw:lce}.
I also check the expansion on an $F_3$ (or {\it fcc}) 
lattice, where most previous data is for the Ising model.
For $\chi_2$, I find agreement up to $L=12$ with Gaunt, Sykes et. al 
\cite{gaunt}.
For both, $\chi_2$ and $\mu_2$,
I agree with the data in the Baker-Kincaid tables \cite{baker}
for the $N=1$ case up to 10th order.
My results are also in agreement with the data by Moore for the 
Ising model on an $F_4$ (or {\it hfcc}) lattice up to 10th order \cite{moore}.
For $\mu_2$, however, I disagree at 10th order in the 14th digit.
Since my $d=3$ results for $\mu_2$ agree with Baker-Kincaid at that order,
and the $d=4$ case is just a trivial extension, I am
confident that my results are correct. 

\btb[p]
\caption[]
{Coefficients $\chi_2^{(L)}$
for $d=4$, $\lambda=\infty$ and $N=4$ on an $F_4$
lattice.}
\bigskip\centering{
\begin{tabular}{rrcl}
\hline \hline
$L$ & \multicolumn{1}{c}{$\chi_2^{(L)}$} & & \\
\hline 
           0&   1  &/&     4     \\
           1&   3     &/&  2       \\
           2&   69    &/&   4       \\
           3&   4 683     &/&  16       \\
           4&   104 997      &/& 16       \\
           5&   11 706 495    &/&   64       \\
           6&   780 269 025      &/& 128       \\
           7&   121 035 719 763     &/&  512       \\
           8&   2 676 992 132 031     &/&  256       \\
           9&   532 080 140 502 519     &/&  1 024       \\
          10&   58 684 317 905 928 645    &/&   2 048       \\
          11&   14 225 715 177 264 006 075   &/&    8 192       \\
          12&   939 747 792 650 917 978 629    &/&   8 192  \\
	  13&	268 832 928 296 330 696 560 119   &/&  32 768 \\
\hline \hline
\end{tabular}}
\etb

\btb[p]
\caption[]
{Coefficients $\mu_2^{(L)}$
for $d=4$, $\lambda=\infty$ and $N=4$ on an $F_4$
lattice.}
\bigskip\centering{
\begin{tabular}{rrcl}
\hline \hline 
$L$ & \multicolumn{1}{c}{$\mu_2^{(L)}$} & & \\
\hline 
           0&  0 &&  \\
           1&   3  &/&     2       \\
           2&   36    &/&   1         \\
           3&   15 123    &/&   16       \\
           4&   57 933     &/&  2          \\
           5&   65 934 375    &/&   64       \\
           6&   1 341 697 815    &/&   32       \\
           7&   986 124 354 243    &/&   512       \\
           8&   197 363 925 717      &/& 2           \\
           9&   5 716 858 384 704 159     &/&  1 024    \\
          10&   177 046 968 775 629 015    &/&   512     \\
          11&   190 703 768 006 610 184 755   &/&    8 192  \\
          12&   1 733 424 402 960 454 006 143    &/&   1 024  \\
	  13&   4 333 450 646 045 939 997 909 759   &/&  32 768  \\
\hline \hline 
\end{tabular}}
\etb

\btb[p]
\caption[]
{Coefficients 
$\chi_4^{(L)}$ for $d=4$, $\lambda=\infty$ and $N=4$ on an $F_4$
lattice.}
\bigskip\centering{
\begin{tabular}{rrcl}
\hline \hline
$L$ & \multicolumn{1}{c}{$\chi_4^{(L)}$} & & \\
\hline 
           0&  -1   &/&    16       \\
           1&  -3     &/&  2       \\
           2&  -351     &/&  8       \\
           3&  -6 069      &/& 4       \\
           4&  -60 615    &/&   1       \\
           5&  -175 900 185  &/&     64      \\ 
           6&  -17 853 643 803  &/&     128       \\
           7&  -2 005 549 316 799   &/&    256       \\
           8&  -247 046 916 740 785    &/&   512       \\
           9&  -33 114 590 817 471 513    &/&   1 024       \\
          10&  -2 399 323 389 666 667 245    &/&   1 024       \\
          11&  -747 560 932 086 148 219 743    &/&   4 096       \\
          12&  -249 187 027 625 458 317 283 449   &/&    16 384  \\
	  13&  -44 243 685 707 698 463 691 882 177 &/&   32 768  \\
\hline \hline 
\end{tabular}}
\etb

\section{The triviality bound}

From the high-temperature expansion 
the relevant physical quantities are calculated
following L\"uscher and Weisz\cite{lw:trivbound1,lw:trivbound2,lw:trivbound3}. 
Some modifications are necessary for the $F_4$ lattice.
In the following analysis I will only consider the 
physical four component case.
The evaluation of the series is improved by taking into account the
scaling laws quoted in section 2. The results 
from the high-temperature expansion 
in the symmetric phase are being used as initial
data to perform an integration of the RG equations
towards the critical line.  
By matching the scaling behavior of the renormalized coupling,
the integration is then continued into the broken phase up to
$m_R=0.5$.
Finally, the result for $m_R/f_\pi$ is compared
with Monte Carlo data.

\subsection{High-temperature series analysis in the symmetric phase}

Approaching the second order critical line $\kappa_c(\lambda)$, 
the correlation length diverges --- i.e.
$\Lambda/ m_R \rightarrow \infty$.
In this limit, the theory is expected to 
behave like an effective, low-energy continuum theory.
As one wants to obtain an upper bound on $\Lambda/m_R$ 
at fixed renormalized coupling $g_R$
(i.e. fixed physics) one has to minimize the 
function $m_R(\lambda)$ along the lines of constant $g_R$
(the RG trajectories).
Thus, the first step is to compute the 
functions $m_R(\kappa,\lambda)$ and $g_R(\kappa,\lambda)$
from the high-temperature series.

The procedure is as follows. First I will determine the
critical line $\kappa_c(\lambda)$. To identify 
possible additional singularities I apply the Pad\'e approximant method.
(Remember that on the $F_4$ lattice there is no simple
anti-ferromagnetic singularity at $\kappa = - \kappa_c$.)
Closer to the critical line, convergence of the series expansion 
becomes rather slow. However, LW found that the rate of convergence 
is ``acceptable'' as long as the correlation length 
does not exceed the average size 
(as measured in lattice spacings)
of the graphs contributing to the cluster expansion. 
I choose as limit $m_R = 0.20$.
In terms of $\kappa$, this condition is always satisfied 
in the region $\kappa \leq 0.98 \kappa_c$.
To obtain  results in the region above $0.98 \kappa_c$, I integrate the 
RG equations (again at fixed $\lambda$). 
The starting point for the integration is taken at the boundary line 
$\kappa = 0.98 \kappa_c$.
As $g_R$ is driven towards zero at the transition,
the perturbative expansions of the RG functions remain valid --- 
provided the coupling is weak enough at the boundary line.
A crude estimate of the 
applicability of renormalized perturbation theory
is given by the tree-level
unitarity bound on the renormalized coupling. 

At this point let me recall that the results from the high-temperature
expansion are for the susceptibilities $\chi_2$, $\mu_2$
and $\chi_4$ as in eqs. (\ref{eq-chi2})---(\ref{eq-chi4}).
From these, $m_R$, $g_R$, $Z_R$, and
$Z^{\cal O}_R$ can be computed via eqs.
(\ref{eq-mreval})---(\ref{eq-zoreval}).
For the analysis, I switch to a new bare coupling $\bar{\lambda}$
which is defined in terms of the connected moments
$\mo^{\rm conn}_{2k}$ (see eq.~(\ref{eq-conmom}) ) as
\ba
	\label{eq-lambar}
	\bar{\lambda} &=& - (1+\frac{N}{2}) 
	\mo^{\rm conn}_4 \left / \; ( \mo^{\rm conn}_2 )^2 \right. \;\;.
\ea
It runs monotonically from 0 to 1. See table~5 for some values.

\btb[p]
\caption[]
{Values for $\kappa_c$ and $\lambda$ at given $\bar{\lambda}$.
$\kappa_c^P$ is from the Pad\'e approximant, $\kappa_c$ is from
the extrapolation of the ratios. }
\bigskip\centering{
\begin{tabular}{llll}
\hline \hline 
\multicolumn{1}{c}{$\bar{\lambda}$} & 
\multicolumn{1}{c}{$\lambda$} & 
\multicolumn{1}{c}{$\kappa_c^P$} &
\multicolumn{1}{c}{$\kappa_c$} \\
\hline 
 0.00	& 0.00			& 0.041667(1)	& 0.041666(1) \\
 0.01	& 0.1723$\cdot 10^{-2}$	& 0.041994(1)	& 0.041997(2) \\
 0.02	& 0.3566$\cdot 10^{-2}$	& 0.042332(2)	& 0.042339(3) \\
 0.03	& 0.5535$\cdot 10^{-2}$	& 0.042680(2)	& 0.042691(5) \\
 0.04	& 0.7639$\cdot 10^{-2}$	& 0.043041(3)	& 0.043054(6) \\
 0.05	& 0.9889$\cdot 10^{-2}$	& 0.043413(5)	& 0.043429(8) \\
 0.06	& 0.1229$\cdot 10^{-1}$	& 0.043797(5)	& 0.043815(9) \\
 0.07	& 0.1487$\cdot 10^{-1}$	& 0.044193(5)	& 0.04421(1) \\
 0.08	& 0.1762$\cdot 10^{-1}$	& 0.044602(6)	& 0.04462(1) \\
 0.09	& 0.2056$\cdot 10^{-1}$	& 0.045024(7)	& 0.04505(1) \\
 0.10	& 0.2370$\cdot 10^{-1}$	& 0.045459(7)	& 0.04549(1) \\
 0.20	& 0.7041$\cdot 10^{-1}$	& 0.05061(1)	& 0.05064(3) \\
 0.30	& 0.1635		& 0.05728(1)	& 0.05732(4) \\
 0.40	& 0.3451		& 0.06514(1)	& 0.06518(4) \\
 0.50	& 0.6791		& 0.07314(2)	& 0.07317(5) \\
 0.60	& 1.256			& 0.08004(4)	& 0.08005(6) \\
 0.70	& 2.244			& 0.08514(6)	& 0.08517(6) \\
 0.80	& 4.127			& 0.08849(6)	& 0.08851(6) \\
 0.90	& 9.347			& 0.09044(8)	& 0.09045(5) \\
 1.00	& $\infty$		& 0.0916(1)	& 0.09155(5) \\
\hline \hline 
\end{tabular}}
\etb

The convergence properties of the high-temperature
expansion have been found to improve 
by reexpanding the series in powers of a so-called
character variable $v$ \cite{lw:trivbound1}.
It can be expressed as
\be
	v = \frac{ \displaystyle
	\sum_{l=1,3,5,\ldots} \frac{1}{l!} (2\kappa)^l
	\frac{\Gamma(\frac{l+2}{2})}{\Gamma(\frac{N+l+1}{2})} J^2_{N+l}
	}{\displaystyle
	\sum_{l=0,2,4,\ldots} \frac{1}{l!} (2\kappa)^l
	\frac{\Gamma(\frac{l+1}{2})}{\Gamma(\frac{N+l}{2})} J^2_{N+l-1}
	} \;.	\label{eq-vtrafo}
\ee
In the Ising limit this reduces to the 
well known transformation $v=\tanh 2 \kappa$.
$\kappa_c(\lambda)$ is given
by the pole in the series expansion of
the susceptibility $\chi_2$ 
and is extracted using the ratio method.
Because all coefficients in the expansion
$ \chi_2 = \sum_{i=0}^{\infty} \chi^{(i)}_2 v^i $
are positive, the pole closest to the origin must be on the positive
real axis. This implies that the singularity is given by 
$v_c = \lim_{i \rightarrow \infty} r_i $ with
$	r_i = \chi^{(i-1)}_2 /  \chi^{(i)}_2 $.
For $i=13$, the ratios have already converged rather well.

Based on the knowledge about the scaling behavior
of the physical singularity in $\chi_2$,
the extrapolation $i \rightarrow \infty$ can be improved.
This behavior is described by eq.~(\ref{eq-chi2scali}) and
can be simulated by an auxiliary function of the form
\ba
	h(z) &=& (1-z)^{-1} \biggl ( 1
	- f(\lambda) \ln(1-z)
	\biggr)^{-\delta_1 / \beta_1} \\
	&=& \sum_{i=0}^{\infty} h^{(i)}z^i \nn \;,
\label{eq-auxfct}
\ea
where $f(\lambda)$ is given in eq.~(\ref{eq-flambda}).
Taking the ratios $r'_i=h^{(i-1)}/h^{(i)}$, one would expect similar 
convergence properties for $r_i$ and $r'_i$. 
These two sequences are then combined to form a ratio 
$R_i=r_i/r'_i$, which somewhat stabilizes the series 
(especially closer to the non-linear limit).
The value for $v_c$ is obtained by fitting 
the monotonic fall-off in $R_i$ to a $1/i^2$ form.
The error $\Delta v_c$ is taken as $\half | R_{13} - v_c |$.

To study the distribution of singularities
of $\chi_2$ in the complex plane
I use the Pad\'e approximant method.
Suppose the singularity of $f(z)$ can be written in the form
$ f(z) = A \; (z-z_{c})^{\alpha} \; \{ 1 + {\cal O}(z-z_{c}) \}$
for $ z \approx z_{c}$.
Then the logarithmic derivative of $f(z)$
exhibits a simple pole at $z=z_{c}$ which is extracted
using the Pad\'e approximant. The exponent $\alpha$ is given by the
residue at the pole.

Results for $\kappa_c$ from both methods
at selected values of $\bar{\lambda}$ are displayed
in table~5. Notice that the values for $\kappa_c$ from the Pad\'e
approximation are consistently below the result from the
extrapolation of the ratios $R_i$. 
The difference is largest for intermediate $\bar\lambda$
where the first order scaling law improvement may not be adequate.
On the other hand, the Pad\'e method completely ignores the additional 
information about the nature of the physical poles
gained from the scaling laws.
Furthermore, the error estimation for $\kappa_c^P$ is somewhat unclear
and it is therefore difficult to draw definite conclusions.
I will thus consider the results from the ratio method superior
and use them throughout.

At $\lambda = \infty$, the critical point is $\kappa_c=0.09155(5)$
which agrees with previous Monte Carlo data, 
$\kappa_c^{MC} = 0.0917(2)$ \cite{bha:f4_num}.
Except for the pole at $\kappa_c^P$, the Pad\'e approximants only give
spurious zeros or poles far away from the origin.
Almost all of the roots in the polynomial $d(z)$ are on or very close to
the real axis. The value for $\alpha$ from the Pad\'e approximant
is close to -1, as expected.

The calculation of the renormalized quantities 
$m_R$, $g_R$, $Z_R$, and $Z^{\cal O}_R$ is carried out along the 
boundary line $\kappa=0.98 \kappa_c$.
From the scaling laws eqs. (\ref{eq-mrscal})---(\ref{eq-zorscal}),
the expansions can be written as
\ba
	m_R &=& \left(1-\frac{v}{v_c}\right)^{1/2}
	v^{-1/2} \; \hat{m}_R(v) \;, \\
	g_R &=&  v^{-2} \hat{g}_R(v) \;, \\
	Z_R &=&  v^{-1} \hat{Z}_R(v) \;,\\
	Z^{\cal O}_R &=& \hat{Z}^{\cal O}_R(v) \;,
\ea
with the reduced expansion
$\hat{m}_R (v) = \sum_{i=0}^{\infty} \hat{m}^{(i)}_R v^i $
and similarly for the other quantities.
The scaling behavior of the reduced expansions can then
be described by 
\ba
	\hat{m}_R (v) &\propto& \left( \;
	1 - f(\lambda) \ln \left(1-\frac{v}{v_c} \right)
	\right)^{\delta_1/2\beta_1} (1+{\cal O}(g_R\ln g_R)) \;, \\
	\hat{g}_R (v) &\propto& \left( \;
	1 - f(\lambda) \ln \left(1-\frac{v}{v_c} \right)
	\right)^{-1} (1+{\cal O}(g_R\ln g_R)) \;, \\
	\hat{Z}_R (v) &\propto& C (1+{\cal O}(g_R\ln g_R)) \;, \\
	\hat{Z}^{\cal O}_R (v) &\propto& \left( \;
	1 - f(\lambda) \ln \left(1-\frac{v}{v_c} \right)
	\right)^{-\delta_1/\beta_1} (1+{\cal O}(g_R\ln g_R)) \;,
\ea
which includes improvement.
For the purpose of extrapolation and error estimation,
auxiliary functions similar to  eq.~(\ref{eq-auxfct})
are being introduced (except in the case of
$\hat{Z}_R(v)$, which tends to a constant).
From the overlap of the reduced expansion with the auxiliary function
one can extract the extrapolated value as well as a reliable
error estimate. In fact, the proportionality factor
$C =  v^i_c \;{\hat{m}^{(i)}_R }/{h^{(i)}} $
is monotonic and stabilizes for $i\geq 10$. 
With the high-temperature expansion up to
13th order, the remainder of the series can be 
reliably estimated as
\be
	\delta = C \left \{ h(z) - \sum_{i=0}^{L-1}
	h^{(i)} z^i
	\right\}_{z=v/v_c} \;,
\ee
such that the extrapolated estimate for the (reduced) mass is given by
\be
	\hat{m}^*_R (v) = \sum_{i=0}^{L-1} \hat{m}^{(i)}_R v^i + \delta \;.
\ee
From the variation in the overlap I obtain an error on $\delta$ and thus
on the mass $m_R$. The quantities $g_R$ and $Z^{\cal O}_R$ are treated
similarly. 
For $Z_R$, however, I evaluate $\hat{Z}_R (v)$ for
$L,L-1,L-2,\ldots$ and extract the best estimate by fitting the 
monotonic behavior of the truncated series to a quadratic fall-off in $L$.

\btb[p]
\caption[]
{High-temperature expansion results for $m_R$, $g_R$, $Z_R$, 
and $Z_R^{\cal O}$ at $\kappa=0.98\kappa_c$.}
\bigskip\centering{
\begin{tabular}{lllr@{.}lll}
\hline \hline
\multicolumn{1}{c}{$\bar{\lambda}$} & 
\multicolumn{1}{c}{$\kappa$} & 
\multicolumn{1}{c}{$m_R$} & 
\multicolumn{2}{c}{$g_R$} & 
\multicolumn{1}{c}{$Z_R$} & 
\multicolumn{1}{c}{$Z_R^{\cal O}$} \\
\hline 
 0.00 & 0.0408333 & 0.285714(1)& 0&0	     & 4.0816326(3)& 0.0102083(1) \\
 0.01 & 0.0411575 & 0.28442(6) & 0&32854(1)  & 4.049486(1) & 0.010342(1) \\
 0.02 & 0.0414919 & 0.2832(1)  & 0&64820(5)  & 4.016840(5) & 0.010479(1) \\
 0.03 & 0.0418369 & 0.2820(2)  & 0&9596(1)   & 3.98370(1)  & 0.010619(3) \\
 0.04 & 0.0421929 & 0.2808(2)  & 1&2633(3)   & 3.95008(2)  & 0.010763(3) \\
 0.05 & 0.0425601 & 0.2797(3)  & 1&5597(5)   & 3.91597(3)  & 0.010910(4) \\
 0.06 & 0.0429388 & 0.2787(3)  & 1&8492(9)   & 3.88140(4)  & 0.011060(5) \\
 0.07 & 0.0433295 & 0.2776(4)  & 2&132(1)    & 3.84637(5)  & 0.011215(6) \\
 0.08 & 0.0437324 & 0.2766(4)  & 2&409(2)    & 3.81090(7)  & 0.011373(7) \\
 0.09 & 0.0441479 & 0.2756(4)  & 2&680(2)    & 3.77500(8)  & 0.011535(8) \\
 0.10 & 0.0445763 & 0.2747(5)  & 2&945(3)    & 3.7387(1)   & 0.011701(9) \\
 0.20 & 0.0496317 & 0.2664(7)  & 5&33(1)     & 3.3573(3)   & 0.01362(2) \\
 0.30 & 0.0561757 & 0.2595(8)  & 7&34(2)     & 2.9653(6)   & 0.01608(3) \\
 0.40 & 0.0638780 & 0.2535(8)  & 9&05(3)     & 2.6068(9)   & 0.01904(4) \\
 0.50 & 0.0717026 & 0.2480(7)  & 10&54(4)    & 2.321(1)    & 0.02221(5) \\
 0.60 & 0.0784481 & 0.2429(7)  & 11&84(3)    & 2.120(1)    & 0.02523(6) \\
 0.70 & 0.0834642 & 0.2381(6)  & 12&98(2)    & 1.992(2)    & 0.02784(6) \\
 0.80 & 0.0867365 & 0.2335(5)  & 14&00(1)    & 1.915(2)    & 0.02999(6) \\
 0.90 & 0.0886387 & 0.2291(4)  & 14&92(3)    & 1.872(2)    & 0.03175(5) \\
 1.00 & 0.0897232 & 0.2248(2)  & 15&8(2)     & 1.848(3)    & 0.03331(3) \\
\hline \hline 
\end{tabular}}
\etb

Results for  $m_R$, $g_R$, $Z_R$, and $Z^{\cal O}_R$ along the
boundary line $\kappa=0.98\kappa_c$ are given in table~6.
All errors are within 2\% and the maximal correlation length is less than
5, compatible with the demand that it be less than the average extent of the
graphs in the linked cluster expansion.
As expected, the maximal value for the coupling $g_R$ along the boundary
line occurs in the $\sigma$-model
limit with $g_R=15.8(2)$, about half the unitarity bound. 

I compare the results from the 13th order calculation 
with those from the 12th order to check
for inconsistencies in the error estimates. At 12th order, the
critical line is shifted to slightly higher $\kappa$ values, but
always much less than one standard deviation. 
The change in $\kappa_c$ has a negligible effect on
$m_R$, $g_r$ and $Z^{\cal O}_R$, the effect on $Z_R$, however, is
noticeable, particularly for $\bar{\lambda \leq 0.1}$.
This indicates that the error for $Z_R$ may be underestimated,
but this will have no effect on the final result for the 
triviality bound.

As a further check, I evaluate the high-temperature series at $\kappa=0.09$
in the $\sigma$-model limit (even though this point is slightly above
the boundary line).
Here, Monte Carlo data \cite{bha:f4_num}
predict $g_R=15.5(1.5)$ and $m_R=0.2075(15)$.
I obtain $g_R=15.2(3)$ and $m_R=0.2054(2)$.
The Monte Carlo mass is significantly heavier, maybe
indicating slight contamination by excited states.

\btb[p]
\caption[]
{Results in the symmetric phase at given $m_R$ for
$\bar{\lambda}= 0.01$. The values for 
$\kappa \leq 0.98 \kappa_c(\bar{\lambda})$
are from the high-temperature expansion, while those for
$\kappa > 0.98 \kappa_c(\bar{\lambda})$
are from the integration of the RG equations.}
\bigskip\centering{
\begin{tabular}{llllll}
\hline \hline
\multicolumn{1}{c}{$\bar{\lambda}$} & 
\multicolumn{1}{c}{$m_R$} & 
\multicolumn{1}{c}{$g_R$} & 
\multicolumn{1}{c}{$Z_R$} & 
\multicolumn{1}{c}{$Z_R^{\cal O}$} & 
\multicolumn{1}{c}{$\kappa$} \\
\hline 
0.01    & 1.00 & 0.4978807(2) & 4.9641184(1) & 0.00840706(1) & 0.03357427(5) \\
	& 0.90 & 0.4604509(4) & 4.7752191(2) & 0.00874153(3) & 0.03490241(8) \\
	& 0.80 & 0.4281371(7) & 4.6061668(3) & 0.00906474(5) & 0.0361834(1) \\
	& 0.70 & 0.400522(1) & 4.4569599(4) & 0.00937124(9) & 0.0373947(2) \\
	& 0.60 & 0.377231(2) & 4.3275968(6) & 0.0096553(2) & 0.0385125(3) \\
	& 0.50 & 0.357933(3) & 4.2180769(8) & 0.0099111(3) & 0.0395125(3) \\
	& 0.40 & 0.342329(5) & 4.128401(1) & 0.0101333(5) & 0.0403707(4) \\
	& 0.30 & 0.330145(8) & 4.058578(1) & 0.0103173(8) & 0.0410653(4) \\
	& 0.20 & 0.321093(9) & 4.008676(1) & 0.0104605(9) & 0.041576(2) \\
	& 0.10 & 0.314617(9) & 3.978685(1) & 0.0105666(9) & 0.041890(2) \\
	& 0.09 & 0.314058(8) & 3.976782(1) & 0.0105760(9) & 0.041910(2) \\
	& 0.08 & 0.313500(8) & 3.975078(1) & 0.0105853(9) & 0.041928(2) \\
	& 0.07 & 0.312937(8) & 3.973574(1) & 0.0105948(9) & 0.041944(2) \\
	& 0.06 & 0.312355(8) & 3.972270(1) & 0.0106046(9) & 0.041957(2) \\
	& 0.05 & 0.311736(8) & 3.971165(1) & 0.0106151(9) & 0.041969(2) \\
	& 0.04 & 0.311051(8) & 3.970260(1) & 0.0106267(9) & 0.041979(2) \\
	& 0.03 & 0.310243(8) & 3.969555(1) & 0.0106405(9) & 0.041986(2) \\
	& 0.02 & 0.309185(8) & 3.969050(1) & 0.0106587(9) & 0.041991(2) \\
	& 0.01 & 0.307477(8) & 3.968744(1) & 0.0106884(9) & 0.041995(2)  \\
\hline \hline
\end{tabular}}
\etb

\btb[p]
\caption[]
{Same as table 7 but for $\bar{\lambda}= 0.1$.}
\bigskip\centering{
\begin{tabular}{llllll}
\hline \hline
\multicolumn{1}{c}{$\bar{\lambda}$} & 
\multicolumn{1}{c}{$m_R$} & 
\multicolumn{1}{c}{$g_R$} & 
\multicolumn{1}{c}{$Z_R$} & 
\multicolumn{1}{c}{$Z_R^{\cal O}$} & 
\multicolumn{1}{c}{$\kappa$} \\
\hline 
0.10    & 1.00 & 4.81569(7) & 4.61008(1) & 0.0091809(1) & 0.0361520(4) \\
	& 0.90 & 4.4304(1) & 4.43251(2) & 0.0095684(3) & 0.0376002(6) \\
	& 0.80 & 4.0942(2) & 4.27330(2) & 0.0099494(5) & 0.039001(1) \\
	& 0.70 & 3.8021(4) & 4.13246(4) & 0.0103196(9) & 0.040330(1) \\
	& 0.60 & 3.5497(6) & 4.00996(5) & 0.010675(2) & 0.041562(2) \\
	& 0.50 & 3.332(1) & 3.90583(6) & 0.011011(3) & 0.042669(3) \\
	& 0.40 & 3.146(2) & 3.82007(8) & 0.011328(5) & 0.043627(3) \\
	& 0.30 & 2.983(3) & 3.7528(1) & 0.011627(8) & 0.044409(3) \\
	& 0.20 & 2.833(3) & 3.7043(1) & 0.011926(9) & 0.04499(1) \\
	& 0.10 & 2.666(3) & 3.67481(9) & 0.01229(1) & 0.04535(1) \\
	& 0.09 & 2.645(2) & 3.67290(9) & 0.01234(1) & 0.04537(1) \\
	& 0.08 & 2.623(2) & 3.67117(9) & 0.01239(1) & 0.04539(1) \\
	& 0.07 & 2.599(2) & 3.66963(9) & 0.01245(1) & 0.04541(1) \\
	& 0.06 & 2.572(2) & 3.66829(9) & 0.01251(1) & 0.04543(1) \\
	& 0.05 & 2.541(2) & 3.66712(9) & 0.01259(1) & 0.04544(1) \\
	& 0.04 & 2.505(2) & 3.66615(9) & 0.01268(1) & 0.04545(1) \\
	& 0.03 & 2.461(2) & 3.66535(9) & 0.01280(1) & 0.04546(1) \\
	& 0.02 & 2.402(2) & 3.66472(9) & 0.01296(1) & 0.04547(1) \\
	& 0.01 & 2.307(2) & 3.66421(9) & 0.01322(1) & 0.04547(1) \\
\hline \hline
\end{tabular}}
\etb

\btb[p]
\caption[]
{Same as table 7 but for $\bar{\lambda}= 1.0$.}
\bigskip\centering{
\begin{tabular}{llr@{.}llll}
\hline \hline
\multicolumn{1}{c}{$\bar{\lambda}$} & 
\multicolumn{1}{c}{$m_R$} & 
\multicolumn{2}{c}{$g_R$} & 
\multicolumn{1}{c}{$Z_R$} & 
\multicolumn{1}{c}{$Z_R^{\cal O}$} & 
\multicolumn{1}{c}{$\kappa$} \\
\hline 
1.00    & 1.00 & 40&241(3) & 2.3957(2) & 0.0200322(3) & 0.0694724(2) \\
	& 0.90 & 36&014(6) & 2.2927(3) & 0.0212659(5) & 0.0725736(4) \\
	& 0.80 & 32&22(1) & 2.1991(5) & 0.022583(1) & 0.0756370(7) \\
	& 0.70 & 28&81(2) & 2.1149(7) & 0.023996(2) & 0.078616(1) \\
	& 0.60 & 25&72(3) & 2.040(1) & 0.025525(4) & 0.081454(2) \\
	& 0.50 & 22&88(6) & 1.975(1) & 0.027200(7) & 0.084092(2) \\
	& 0.40 & 20&2(1) & 1.920(2) & 0.02908(1) & 0.086459(3) \\
	& 0.30 & 17&7(2) & 1.875(2) & 0.03129(2) & 0.088481(4) \\
	& 0.20 & 15&1(3) & 1.841(3) & 0.03408(4) & 0.090073(3) \\
	& 0.10 & 12&1(1) & 1.819(3) & 0.0383(1) & 0.09113(5) \\
	& 0.09 & 11&7(1) & 1.818(3) & 0.0389(1) & 0.09121(5) \\
	& 0.08 & 11&3(1) & 1.816(3) & 0.0395(1) & 0.09127(5) \\
	& 0.07 & 11&0(1) & 1.815(3) & 0.0402(1) & 0.09133(5) \\
	& 0.06 & 10&5(1) & 1.814(3) & 0.0411(1) & 0.09139(5) \\
	& 0.05 & 10&1(1) & 1.812(3) & 0.0420(1) & 0.09143(5) \\
	& 0.04 & 9&59(9) & 1.811(3) & 0.0432(2) & 0.09147(5) \\
	& 0.03 & 9&02(8) & 1.810(3) & 0.0446(2) & 0.09150(5) \\
	& 0.02 & 8&31(7) & 1.809(2) & 0.0466(2) & 0.09152(5) \\
	& 0.01 & 7&33(5) & 1.808(2) & 0.0498(3) & 0.09154(5) \\
\hline \hline 
\end{tabular}}
\etb

I now turn to the integration of the RG equations
(\ref{eq-beta})--(\ref{eq-kappa}) 
in the region $0.98\kappa_c \leq \kappa \leq \kappa_c$
which is done numerically, starting from the 
results of the high-temperature series at the boundary line.
The results are displayed in tables 7---9 for selected values of
$\bar{\lambda}$ and in the range $0.01 \leq m_R \leq 1.0$.
The values for $\kappa \leq 0.98\kappa_c$ are from the
high-temperature series analysis. 
For the integration I use the RG functions up to three loops 
including the lattice dependent scaling violating terms up to 1-loop order. 
At this order, the scaling violations in the
region $\kappa > 0.98 \kappa_c$ are always less than 6\%
of the universal part. Therefore, the mass dependent terms are
omitted at higher orders.
From the tables notice that $\kappa \rightarrow
\kappa_c$ as $m_R \rightarrow 0$, as it should be.
Errors are estimated by propagating the 
errors of the initial data and do not include the 
systematic uncertainties due to the approximation
of the RG functions. 
To estimate this systematics, I use the  2-loop approximations instead,
and the effect for $g_R$ in the non-linear limit
is demonstrated in fig.~2. 
Generally, the difference between the 
2-loop and 3-loop approximations is maximal in the non-linear limit.

\bfg[t]
	  \vspace{8cm}  
	  \includegraphics{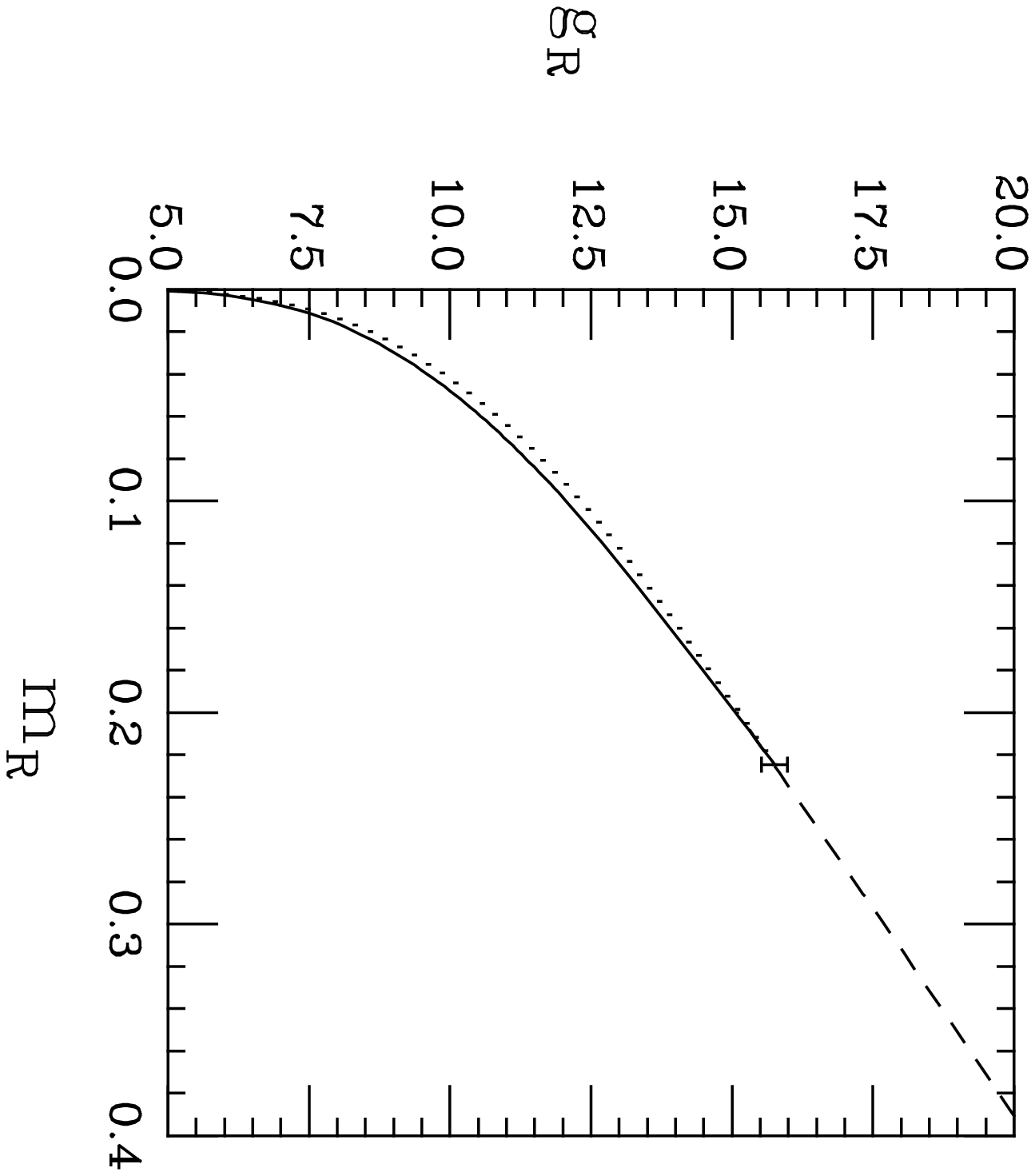}
\caption[]
{Comparison of the solutions of the RG
equations at two and three loops for $g_R$ at $\bar{\lambda}=1$.
The dashed line is the result from the high-temperature expansion,
while the solid line is the continuation obtained by integrating the
3-loop approximation of the $\beta$-funxction.
The dotted curve corresponds to the 2-loop approximation.
It is evident that the deviations in the symmetric phase are of the
same size as the 
series expansion error in $g_R$ calculated at the boundary line.}
\efg

\subsection{Continuation of the analysis to the broken symmetry phase}

As indicated in section 2, to continue
the solution of the theory into the broken phase 
one needs to evaluate the
integration constants $C_1$, $C_2$, and $C_3$ at the transition. 
These constants are defined through the scaling behavior of
$m_R$, $Z_R$, and $Z^{\cal O}_R$ in 
eqs.~(\ref{eq-mr:scaling}), (\ref{eq-zr:scaling}) 
and (\ref{eq-zor:scaling}) and
can be computed by numerically integrating the RG
equations down to small $m_R$,
for instance $m_R=10^{-8}$, using the 3-loop expressions.
In this region $\ln C_i$ is linear in $g_R$
and the limit $g_R \ra 0$ can be extracted.
The results for some values of $\bar{\lambda}$ are presented in table~10.

\btb[p]
\caption[]
{Values for $\ln C_i$ at the critical line.}
\bigskip\centering{
\begin{tabular}{lr@{.}lll}
\hline \hline
\multicolumn{1}{c}{$\bar{\lambda}$} & 
\multicolumn{2}{c}{$\ln C_1$} & 
\multicolumn{1}{c}{$\ln C_2$} &
\multicolumn{1}{c}{$\ln C_3$} \\
\hline 
 0.01 & 126&429(4) & 1.3782607(3) & -5.1270(1) \\
 0.02 &  64&356(5) & 1.370040(1) & -4.7730(2) \\
 0.03 &  43&563(7) & 1.361631(2) & -4.5625(3) \\
 0.04 &  33&116(8) & 1.353031(4) & -4.4106(5) \\
 0.05 &  26&82(1)  & 1.344237(6) & -4.2907(6) \\
 0.06 &  22&60(1)  & 1.335249(9) & -4.1909(7) \\
 0.07 &  19&57(1)  & 1.32606(1)  & -4.1049(9) \\
 0.08 &  17&29(1)  & 1.31668(1)  & -4.029(1) \\
 0.09 &  15&51(1)  & 1.30710(2)  & -3.961(1) \\
 0.10 &  14&07(2)  & 1.29731(2)  & -3.898(1) \\
 0.20 &   7&50(2)  & 1.18865(8)  & -3.441(3) \\
 0.30 &   5&22(2)  & 1.0636(2)   & -3.110(4) \\
 0.40 &   4&04(2)   & 0.9340(3)  & -2.831(4) \\
 0.50 &   3&32(1)   & 0.8173(4)  & -2.596(4) \\
 0.60 &   2&82(1)   & 0.7263(6)  & -2.407(4) \\
 0.70 &   2&460(7)  & 0.6633(8)  & -2.259(3) \\
 0.80 &   2&184(4)  & 0.624(1)   & -2.144(3) \\
 0.90 &   1&963(7)  & 0.601(1)   & -2.052(3) \\
 1.00 &   1&78(4)   & 0.588(1)	 & -1.975(9) \\
\hline \hline 
\end{tabular}}
\etb

Given the integration constants on the high-temperature side
of the transition,  
the constants $C'_i$ in the broken phase are fixed through
eq.~(\ref{eq:match}).
$m_R$, $Z_R$, and $Z^{\cal O}_R$
can then be calculated using the broken phase scaling laws.
I choose $g_R=10^{-8}$ --- small enough for order $g_R^2$ 
corrections to be negligible ---
and continue integrating the RG equations up to $m_R=0.5$,
again using the 3-loop expressions for the RG functions
including scaling violating terms up to 1-loop.
The results at $m_R=0.5$ 
are displayed in table~11 for selected values of $\bar{\lambda}$.
Results along the lines of constant $\bar{\lambda}$ are shown
in tables 12---14.
Errors are estimated by propagating the uncertainties in the
integration constants only.
The renormalized coupling is maximal in the non-linear limit where it
reaches $g_R = 18(1)$ at $m_R=0.5$, about 2/3 the unitarity bound.
The assumption of applicability of perturbative RG
is justified by comparing the 3-loop with the 2-loop results:
the changes in the region $m_R \leq 0.5$ are not unreasonable.

In fig.~3 I compare the results for $m_R/f_\pi$ 
with recent Monte Carlo data \cite{bha:f4_num}.
Both methods are in fair agreement for $m_R < 0.5$.
For larger masses, however,
one leaves the perturbative regime and the integration
of the perturbative RG functions cannot be trusted anymore. 
Also, the $L\rightarrow \infty$ extrapolation of the Monte Carlo data
may be spoiled by $\sigma$-particle decay, perhaps leading to the
systematic difference evident in fig.~3.
At $\Lambda/m_R=2$ I obtain for an upper bound on the Higgs mass
\be
	m_R/f_\pi \leq 2.46 \pm 0.02_{\rm HTE} \pm 0.08_{\rm PT} \;,
\ee
where the first error is from the extrapolation of the 
high-temperature expansion and the second error is from the perturbative
expansion of the RG functions. The latter 
is estimated by comparing the 2-loop and the 3-loop results.
The above numbers translate to about 
$ 600 \pm 5_{\rm HTE} \pm 20_{\rm PT}$~GeV.
The Monte Carlo result is 2.3(2) \cite{bha:f4_num},
somewhat below the number from the high-temperature expansion.
The agreement is better
in the second reference of \cite{bha:f4_num},
where in a similar plot the RG integration is based on Monte Carlo 
data in the symmetric phase.
This indicates that the systematic difference 
in fig.~3 is mainly due to the expansion itself and not
to the perturbative RG integration.

The $F_4$ results compare to $m_R/f_\pi \leq 2.6(3)$ 
(or 640(70)~GeV) from the high-temperature expansion on a hypercubic lattice 
\cite{lw:strong,lw:trivbound3}.
Here Monte Carlo simulations give 2.7(1) \cite{triv:num}, slightly
{\em above} the semi-analytical result.
Interestingly, the hypercubic results are in reverse order.

\btb[p]
\caption[]
{Results in the broken phase at  $m_R=0.5$ from 
integrating the RG equations.}
\bigskip\centering{
\begin{tabular}{llr@{.}lll}
\hline \hline
\multicolumn{1}{c}{$\bar{\lambda}$} & 
\multicolumn{1}{c}{$\kappa$} &
\multicolumn{2}{c}{$g_R$} & 
\multicolumn{1}{c}{$Z_R$} & 
\multicolumn{1}{c}{$Z_R^{\cal O}$} \\
\hline 
 0.01 & 0.043384(1) & 0&290944(8) & 3.841565(1) & 0.010567(1) \\
 0.02 & 0.043739(3) & 0&57660(5) & 3.810283(4) & 0.010709(3) \\
 0.03 & 0.044106(4) & 0&8573(1) & 3.778535(9) & 0.010855(4) \\
 0.04 & 0.044484(5) & 1&1335(3) & 3.74633(2) & 0.011004(6) \\
 0.05 & 0.044874(6) & 1&4053(5) & 3.71368(2) & 0.011156(8) \\
 0.06 & 0.045276(8) & 1&6731(8) & 3.68059(3) & 0.01131(1) \\
 0.07 & 0.045691(9) & 1&937(1) & 3.64707(4) & 0.01147(1) \\
 0.08 & 0.04612(1)  & 2&197(2) & 3.61313(6) & 0.01164(2) \\
 0.09 & 0.04656(1)  & 2&454(2) & 3.57879(7) & 0.01180(2) \\
 0.10 & 0.04701(1)  & 2&708(3) & 3.54407(8) & 0.01198(2) \\
 0.20 & 0.05238(2)  & 5&08(1)  & 3.1800(3) & 0.01397(5) \\
 0.30 & 0.05933(2)  & 7&21(3)  & 2.8068(5) & 0.01652(9) \\
 0.40 & 0.06752(3)  & 9&15(4)  & 2.4659(8) & 0.0196(1) \\
 0.50 & 0.07585(3)  & 10&92(4) & 2.195(1) & 0.0229(1) \\
 0.60 & 0.08305(4)  & 12&55(4) & 2.004(1) & 0.0261(1) \\
 0.70 & 0.08844(4)  & 14&05(3) & 1.882(2) & 0.0288(1) \\
 0.80 & 0.09199(5)  & 15&46(2) & 1.809(2) & 0.0310(1) \\
 0.90 & 0.09409(4)  & 16&79(4) & 1.768(2) & 0.0329(1) \\
 1.00 & 0.09534(2)  & 18&1(3)  & 1.745(2) & 0.0345(6) \\
\hline \hline 
\end{tabular}}
\etb

\btb[p]
\caption[]
{Results in the broken phase at given $m_R$ for $\bar{\lambda}= 0.01$.}
\bigskip\centering{
\begin{tabular}{llllll}
\hline \hline
\multicolumn{1}{c}{$\bar{\lambda}$} & 
\multicolumn{1}{c}{$m_R$} & 
\multicolumn{1}{c}{$g_R$} & 
\multicolumn{1}{c}{$Z_R$} & 
\multicolumn{1}{c}{$Z_R^{\cal O}$} & 
\multicolumn{1}{c}{$\kappa$} \\
\hline 
0.01  & 0.01 & 0.306186(8) & 3.967302(1) & 0.010722(1) & 0.041998(2) \\
      & 0.02 & 0.307806(8) & 3.967149(1) & 0.010693(1) & 0.042000(2) \\
      & 0.03 & 0.308731(8) & 3.966898(1) & 0.010676(1) & 0.042002(2) \\
      & 0.04 & 0.309357(8) & 3.966549(1) & 0.010664(1) & 0.042006(2) \\
      & 0.05 & 0.309810(8) & 3.966101(1) & 0.010655(1) & 0.042011(2) \\
      & 0.06 & 0.310147(8) & 3.965555(1) & 0.010647(1) & 0.042017(2) \\
      & 0.07 & 0.310396(8) & 3.964909(1) & 0.010641(1) & 0.042023(2) \\
      & 0.08 & 0.310576(8) & 3.964165(1) & 0.010635(1) & 0.042031(2) \\
      & 0.09 & 0.310698(8) & 3.963322(1) & 0.010630(1) & 0.042040(2) \\
      & 0.10 & 0.310770(8) & 3.962380(1) & 0.010626(1) & 0.042051(2) \\
      & 0.20 & 0.309520(8) & 3.947501(1) & 0.010597(1) & 0.042211(2) \\
      & 0.30 & 0.305563(8) & 3.922586(1) & 0.010580(1) & 0.042481(2) \\
      & 0.40 & 0.299318(8) & 3.887393(1) & 0.010570(1) & 0.042869(2) \\
      & 0.50 & 0.290944(8) & 3.841565(1) & 0.010567(1) & 0.043384(1) \\
\hline \hline 
\end{tabular}}
\etb

\btb[p]
\caption[]
{Same as table 12 but for $\bar{\lambda}= 0.1$.}
\bigskip\centering{
\begin{tabular}{llllll}
\hline \hline
\multicolumn{1}{c}{$\bar{\lambda}$} & 
\multicolumn{1}{c}{$m_R$} & 
\multicolumn{1}{c}{$g_R$} & 
\multicolumn{1}{c}{$Z_R$} & 
\multicolumn{1}{c}{$Z_R^{\cal O}$} & 
\multicolumn{1}{c}{$\kappa$} \\
\hline 
0.10  & 0.01 & 2.237(2) & 3.65503(9) & 0.01354(2) & 0.04549(1) \\
      & 0.02 & 2.325(2) & 3.65471(9) & 0.01328(2) & 0.04549(1) \\
      & 0.03 & 2.380(2) & 3.65437(9) & 0.01312(2) & 0.04549(1) \\
      & 0.04 & 2.420(2) & 3.65397(9) & 0.01301(2) & 0.04550(1) \\
      & 0.05 & 2.453(2) & 3.65349(9) & 0.01293(2) & 0.04550(1) \\
      & 0.06 & 2.479(2) & 3.65293(9) & 0.01286(2) & 0.04551(1) \\ 
      & 0.07 & 2.502(2) & 3.65230(9) & 0.01279(2) & 0.04552(1) \\
      & 0.08 & 2.522(2) & 3.65158(9) & 0.01274(2) & 0.04553(1) \\
      & 0.09 & 2.540(2) & 3.65078(9) & 0.01269(2) & 0.04554(1) \\
      & 0.10 & 2.556(2) & 3.64989(9) & 0.01265(2) & 0.04555(1) \\
      & 0.20 & 2.658(3) & 3.63640(9) & 0.01237(2) & 0.04573(1) \\
      & 0.30 & 2.705(3) & 3.61440(9) & 0.01220(2) & 0.04603(1) \\ 
      & 0.40 & 2.719(3) & 3.58373(9) & 0.01207(2) & 0.04646(1) \\
      & 0.50 & 2.708(3) & 3.54407(8) & 0.01198(2) & 0.04701(1) \\
\hline \hline 
\end{tabular}}
\etb

\btb[p]
\caption[]
{Same as table 12 but for $\bar{\lambda}= 1.0$.}
\bigskip\centering{
\begin{tabular}{llr@{.}llll}
\hline \hline
\multicolumn{1}{c}{$\bar{\lambda}$} & 
\multicolumn{1}{c}{$m_R$} & 
\multicolumn{2}{c}{$g_R$} & 
\multicolumn{1}{c}{$Z_R$} & 
\multicolumn{1}{c}{$Z_R^{\cal O}$} & 
\multicolumn{1}{c}{$\kappa$} \\
\hline 
1.00  & 0.01 & 6&64(4) & 1.793(2) & 0.0538(7) & 0.09156(5) \\
      & 0.02 & 7&43(5) & 1.792(2) & 0.0509(6) & 0.09156(5) \\
      & 0.03 & 7&99(6) & 1.792(2) & 0.0491(6) & 0.09158(5) \\
      & 0.04 & 8&43(7) & 1.791(2) & 0.0478(6) & 0.09159(5) \\
      & 0.05 & 8&81(8) & 1.790(2) & 0.0468(6) & 0.09161(5) \\
      & 0.06 & 9&14(8) & 1.790(2) & 0.0460(6) & 0.09163(5) \\
      & 0.07 & 9&45(9) & 1.789(2) & 0.0452(6) & 0.09166(5) \\
      & 0.08 & 9&73(9) & 1.789(2) & 0.0446(6) & 0.09169(5) \\
      & 0.09 & 10&0(1) & 1.788(2) & 0.0440(6) & 0.09172(5) \\
      & 0.10 & 10&2(1) & 1.787(2) & 0.0435(6) & 0.09176(5) \\
      & 0.20 & 12&3(1) & 1.780(2) & 0.0400(6) & 0.09230(4) \\
      & 0.30 & 14&1(2) & 1.770(2) & 0.0377(6) & 0.09312(2) \\
      & 0.40 & 16&0(2) & 1.758(2) & 0.0360(6) & 0.09415(2) \\
      & 0.50 & 18&1(3) & 1.745(2) & 0.0345(6) & 0.09534(2) \\
\hline \hline 
\end{tabular}}
\etb

\bfg[t]
	  \vspace{8cm}  
	  \includegraphics{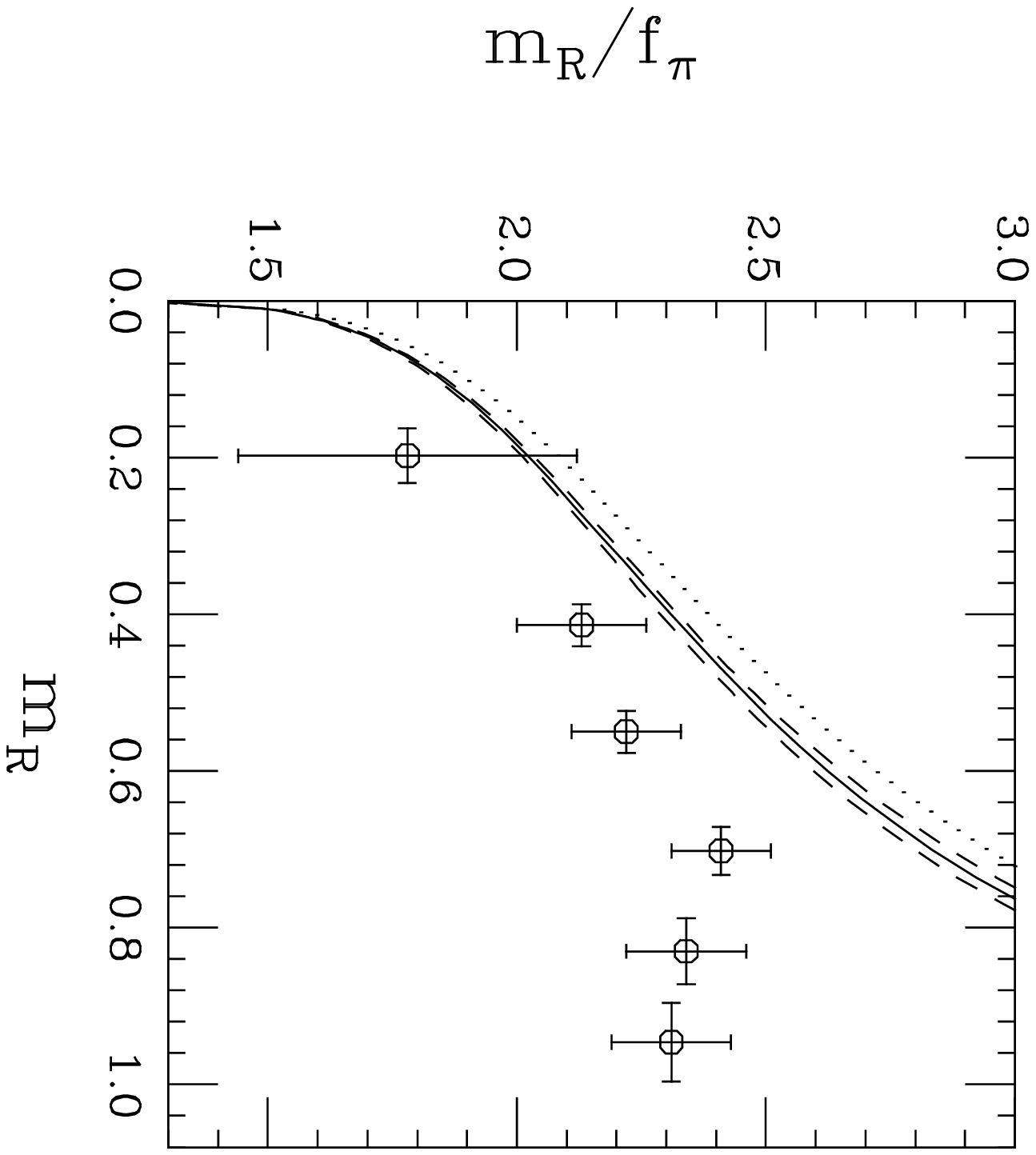}
\caption[]
{Ratio $m_R/f_\pi$ versus $m_R$ in lattice units. The solid line is from 
integrating the RG equations. The dashed line gives the propagated
error and the dotted line is the 2-loop result.
The data points are from a Monte Carlo simulation \cite{bha:f4_num}.
}
\efg

Even though the maximal cutoff
on the $F_4$ lattice seems to be about 6\% lower 
than LW's hypercubic result \cite{lw:trivbound2}, 
this statement by itself is not meaningful.
One must not directly compare the two cutoffs as they have
different meanings on different lattices. 
This is evident in the lattice dependence of the scaling violations.
Rather, one should consider the cutoff dependence of physical quantities,
such as the pion---pion 90 degree scattering cross section.
By demanding that the deviation 
\be
\delta = \frac{\left( d \sigma / d \theta \right)^{\rm lattice}}
	      {\left( d \sigma / d \theta \right)^{\rm continuum}}
	- 1
\ee
be less than (say) 0.3\% for all energies up to some center-of-mass energy,
for instance up to $2m_R$, one defines a physically relevant bound.
On the $F_4$ lattice, $\delta < 0.3\% $ translates to $m_R \approx 0.5$, 
whereas on the hypercubic lattice the same requirement gives 
$m_R \approx 0.22$ \cite{bha:f4_anl}.
From \cite{lw:strong}, this value corresponds to 
$m_R/f_\pi=2.2$ or roughly 540~GeV, which is actually
about 10\% below the $F_4$ bound.

To summarize, in the broken phase I find that 
for $\Lambda/m_R \geq 2$ the coupling never exceeds 2/3 of the 
triviality bound (as in the hypercubic case, see \cite{lw:trivbound3}),
indicating that there is in fact not much room for a heavy Higgs ---
if one insists on keeping the naive lattice action.
A Monte Carlo study including
dimension six operators parameterizing the leading order 
($\Lambda^{-2}$) cutoff effects
has shown that the bound may be as high as 710(60) GeV\cite{hknv}.
A high-temperature expansion for the improved action case would
be desirable but poses quite a complicated task.

Going from 12th to 13th order in the expansion
quadruples the number of contributing graphs and one reaches the 
limits of todays computational resources.
Furthermore, at this order, the uncertainty due to the 
perturbative RG functions
is by far larger than the uncertainty due to the series expansion.
Hence it does not pay to attempt a 14th order calculation without 
simultaneous improvement in the perturbative results for the 
RG functions. 

Finally, I would like to remind the reader that the solution 
of the $\phi^4$ theory as presented here
rests on the assumption that perturbation theory can be applied for
sufficiently small couplings (thereby implicitly assuming triviality),
and relies on the fact that the perturbative region overlaps with the region
in the symmetric phase where the linked cluster expansion can be
evaluated.
These issues are discussed in detail in LW's papers.

The advantages of the semi-analytical approach 
over Monte Carlo methods are evident:
the lack of finite size effects and the absence of statistical errors.
Furthermore, one does not have to worry about the $\sigma$-particle decay
which complicates the extraction of $m_\sigma$ on larger
lattices in Monte Carlo simulations. 
However, with respect to fig.~3 it is not clear from which source 
the differences arise. All one can say is that
the comparison gives a good estimate of systematic errors which are
very different in both methods and presumably do not alter the results
in the same direction. 

\section*{Acknowledgements}

I would like to thank my former advisor, Urs~M.~Heller for
invaluable support. 
Thanks also go to Pavlos Vranas and Herbert Neuberger.
This project was supported in part by the Department of Energy under 
contract number DE-FC05-85~ER250000 and the Supercomputer Computations 
Research Institute and The Florida State University Department of Physics
through use of their facilities.
The linked cluster expansion was performed on the cluster of IBM RS6000
workstations at SCRI.
The final stage of this project was carried out 
while the author was at Columbia University, supported under contract
number DE-FG02-92~ER40699.

\section*{Appendix}

This appendix gives the (slightly corrected) coefficients 
for the RG function (cf. \cite{lw:trivbound3,bha:f4_anl}).
Writing 
\ba
	\label{eq-bdef}
	\beta(0,g_R) &=& g_R \sum_{\nu=1}^{\infty}
	\beta_{\nu} g^{\nu}_R \;, \\
	\label{eq-gdef}
	\gamma(0,g_R) &=& \sum_{\nu=1}^{\infty}
	\gamma_{\nu} g^{\nu}_R \;, \\
	\label{eq-ddef}
	\delta(0,g_R) &=&  \sum_{\nu=1}^{\infty}
	\delta_{\nu} g^{\nu}_R \;, 
\ea
the universal coefficients are given by
\ba
	\beta_1 &=& \frac{1}{3} (N+8) (16\pi^2)^{-1} \;, \\
	\beta_2 &=& - \frac{1}{3} (3N+14) (16\pi^2)^{-2} \;, \\
	\beta_3 &=& \frac{1}{432} \biggl \{61  N^2 + 1782N +5744 
	+ 192 \zeta(3) (5N+22) \nn \\
	& & - 2k(N+8)(13N+62) \biggr \} (16\pi^2)^{-3} \;, \\
	\gamma_1 &=& 0 \;, \\
	\gamma_2 &=& \frac{1}{36} (N+2) (16\pi^2)^{-2} \;, \\
	\gamma_3 &=& -\frac{1}{432} (N+2)(N+8)(1-2k)(16\pi^2)^{-3} \;, \\
	\delta_1 &=& - \frac{1}{3} (N+2) (16\pi^2)^{-1} \;, \\
	\delta_2 &=& \frac{5}{18} (N+2) (16\pi^2)^{-2} \;, \\
	\delta_3 &=& - \frac{1}{216} (N+2) \biggl \{ 31N +221 
	+6k(N+1)\biggr \} (16\pi^2)^{-3} \;,
\ea
where
\ba
	k &=& 1.62520965 \;, \\
	\zeta(3) &=& 1.20205690 \;,
\ea
the latter being the Riemann zeta function.
The non-universal scaling violating terms 
up to one loop are
\ba
	\beta(m_R,g_R) &=& g_R \sum_{\nu=0}^{\infty}
	u_{\nu}(m_R) g^{\nu}_R \;, \\
	\gamma(m_R,g_R) &=& \sum_{\nu=0}^{\infty}
	v_{\nu}(m_R) g^{\nu}_R \;, \\
	\delta(m_R,g_R) &=&  \sum_{\nu=0}^{\infty}
	w_{\nu}(m_R) g^{\nu}_R \;, 
\ea
with the coefficients
\ba
	u_0 &=& \frac{4m^2_R}{4+m^2_R} \;, \\
	u_1 &=& \frac{1}{6} u_0 \left\{
	(N+8)(4+m^2_R)J_3(m^2_R) -6 J_2(m^2_R) \right. \nn \\
	& &  -(N+2) \frac{1}{4+m^2_R} \left. J_1(m^2_R)	\right\} \;, \\
	v_0 &=& \frac{1}{4} u_0 \;, \\
	v_1 &=& \frac{1}{24}(N+2) u_0 \left\{
	J_2(m^2_R)\right. - \frac{1}{4+m^2_R} \left. J_1(m^2_R)
	\right\} \;, \\
	w_0 &=& - \frac{1}{2} u_0 \;, \\
	w_1 &=& -2v_1 -  \frac{1}{6}(N+2) \left. m^2_R \right\{
	(4+m^2_R)J_3(m^2_R)-2J_2(m^2_R)  \nn \\
	& &  + \frac{1}{4+m^2_R} \left. J_1(m^2_R)
	\right\} \;.
\ea
The lattice integrals $J_p(m^2_R)$ are defined as
\be
	J_p(m^2_R) = \int_k \frac{1}{[g(k)+m^2_R]^p} 
\ee
and are evaluated numerically.

In the broken phase, 
the universal coefficients $\beta_1$, $\beta_2$, $\gamma_1$, and
$\delta_1$ take the
same form as in the symmetric phase; the differing coefficients are
given by 
\ba
	\beta_3 &=& \biggl \{ - \frac{\pi^2}{108}N^3 -0.6038002N^2
	+6.006124N 
	+ 11.10641 \biggr \} (16\pi^2)^{-3} , \\
	\gamma_2 &=& -\frac{1}{6} (16\pi^2)^{-2} \;, \\
	\gamma_3 &=& \biggl \{ -0.0452317896N + 0.211712032
	\biggr\} (16\pi^2)^{-3} \;, \\
	\delta_2 &=& \frac{1}{18} \biggl\{(5+2\sqrt{3}\pi)N+16
	- 2\sqrt{3} \pi \biggr\} (16\pi^2)^{-2} \;, \\
	\delta_3 &=& \biggl\{ \frac{\pi^2}{108}N^3 + 
	1.2932334N^2-4.320590N 
	+ 1.620019 \biggr \} (16\pi^2)^{-3} \;.
\ea
For the function $\eps$, the power series is written as
\ba
	\eps(0,g_R) &=&  \sum_{\nu=0}^{\infty}
	\eps_{\nu} g^{\nu}_R \;, 
\ea
with the universal coefficients
\ba
	\eps_0 &=& 1 \;, \\
	\eps_0 &=& \frac{1}{6} \biggl\{-N+3-2\sqrt{3}\pi
	\biggr\} (16\pi^2)^{-1} \;, \\
	\eps_0 &=& \biggl\{ 0.503662N + 2.487506
	\biggr\} (16\pi^2)^{-2} \;.
\ea
For completeness, I quote 
the scaling violating terms (taken from \cite{bha:f4_anl})
up to terms of  order $m^4_R \ln m^2_R$
\ba
u_0 &=& 2- \frac{m_R \sinh m_R}{2 \sinh^2 \half m_R} 
	-\frac{m_R \sinh m_R}{2 - \sinh^2 \half m_R}  \;, \\
u_1 &=& \beta_1 + m^2_R \biggl\{ -2 \ln m^2_R + N \left(
	\frac{7}{72} + \frac{\pi^2}{3}r_0 -\frac{4\pi^2}{9}r_1
	- \frac{16\pi^2}{9}r_2 	\right) \nn \\
    & &  + \frac{13}{36} - \frac{\pi}{\sqrt{3}} + \frac{2\pi^2}{3}r_0
	- \frac{112\pi^2}{9}r_1 +\frac{704\pi^2}{9}r_2
	\biggr\} (16\pi^2)^{-1} \nn \\
    & & + \,{\cal O} (m^4_R \ln m^2_R) \;, \\
v_0 &=& -\frac{1}{4}\; \frac{m_R \sinh m_R}{2 - \sinh^2 \half m_R}  \;, \\
v_1 &=& m^2_R \biggl\{ \frac{1}{48}(N+2)\ln m^2_R +N \left( 
	\frac{\pi^2}{3}r_0 + \frac{4\pi^2}{3}r_1 \right) 
	 +\frac{2\pi^2}{3}r_0 +\frac{8\pi^2}{9}r_1 \nn \\
    & & -\frac{64\pi^2}{9}r_2 + \frac{3\sqrt{3}\pi -4}{12}
	\biggr\} (16\pi^2)^{-1}  + {\cal O} (m^4_R \ln m^2_R) \;, \\
w_0 &=& \frac{m_R \sinh m_R \sinh^2 \half m_R}{4 - \sinh^4 \half m_R}  \;, \\
w_1 &=& \delta_1 +m^2_R \biggl\{\frac{1}{12}(N+2)\ln m^2_R
	+N\left( \frac{\pi^2}{3}r_0 + \frac{16\pi^2}{9}r_1
	+\frac{16\pi^2}{9}r_2 -\frac{7}{72} \right) \nn \\
     & & + \frac{2\pi^2}{3}r_0 -\frac{32\pi^2}{9}r_1 -\frac{224\pi^2}{9}r_2
	-\frac{5}{18} + \frac{\sqrt{3}\pi}{36}
	\biggr\} (16\pi^2)^{-1}  \nn \\
     & & + \, {\cal O} (m^4_R \ln m^2_R) \;, 
\ea
where the constants $r_0$, $r_1$, $r_2$ are given by
\ba
	\label{eq-r0}
	r_0 &=& 0.13823047 \;, \\
	\label{eq-r1}
	r_1 &=& - 0.04029906 \;, \\
	\label{eq-r2}
	r_2 &=& 0.00763417 \;.
\ea
For $\eps$, the $m_R$-dependent expansion is written as
\ba
	\eps(m_R,g_R) &=&  \sum_{\nu=0}^{\infty}
	x_{\nu}(m_R) g^{\nu}_R \;, 
\ea
with the coefficients
\ba
x_0 &=& \frac{\sinh m_R}{m_R} \;
 	 \frac{2 + \sinh^2 \half m_R}{2 - \sinh^2 \half m_R}  \;, \\
x_1 &=& \eps_1 + m^2_R \biggl\{ \ln m^2_R +N \left( -\frac{1}{16}
	- \frac{\pi^2}{6}r_0 + \frac{2\pi^2}{9}r_1 +\frac{8\pi^2}{9}r_2
	\right) \nn \\
    & & - \frac{\pi^2}{3}r_0 +\frac{20\pi^2}{3}r_1 -\frac{112\pi^2}{3}r_2
	-\frac{11}{24} + \frac{7\sqrt{3}\pi}{36}
	\biggr\} (16\pi^2)^{-1}  \nn \\
    & & + {\cal O} (m^4_R \ln m^2_R) \;.
\ea

\ed